\documentclass[aps,twocolumn,showpacs,groupedaddress,floatfix,footinbib]{revtex4-1}
\usepackage{graphicx}
\usepackage{dcolumn}
\usepackage{bm}				
\usepackage{amsfonts}
\usepackage{amssymb}
\usepackage{amsmath}
\usepackage{ulem} 
\usepackage{graphicx}		
\usepackage{xcolor, soul}	
\usepackage{verbatim}
\usepackage[T1]{fontenc}
\usepackage{pifont}
\usepackage{appendix}
\usepackage{multirow}
\usepackage{notes2bib}
\usepackage{physics}
\usepackage{hyperref}

\begin{document}


\title{Quantum geometric bound and ideal condition for Euler band topology}

\author{Soonhyun Kwon$^{1,2}$}
\author{Bohm-Jung Yang$^{1,2,3}$}
\email{bjyang@snu.ac.kr}
\affiliation{
	$^1$Department of Physics and Astronomy, Seoul National University, Seoul 08826, Korea\\
	$^{2}$Center for Theoretical Physics (CTP), Seoul National University, Seoul 08826, Korea\\
	$^3$Center for Correlated Electron Systems, Institute for Basic Science (IBS), Seoul 08826, Korea
}

\date{\today}

\begin{abstract}
Understanding the relationship between quantum geometry and topological invariants is a central problem in the study of topological states. 
In this work, we establish the relationship between the quantum metric and the Euler curvature in two-dimensional systems with space-time inversion $I_{ST}$ symmetry satisfying $I^2_{ST}=+1$.
As $I_{ST}$ symmetry imposes the reality of the wave function with vanishing Berry curvature, the well-known inequality between the quantum metric and the Berry curvature is not meaningful in this class of systems.
We find that the non-abelian quantum geometric tensor of two real bands exhibits an intriguing inequality between the off-diagonal Berry curvature and the quantum metric, which in turn gives the inequality between the quantum volume and the Euler invariant.
Moreover, we show that the saturation condition of the inequality is deeply related to the ideal condition for Euler bands, which provides a criterion for the stability of fractional topological phases in interacting Euler bands.
Our findings demonstrate the potential of the quantum geometry as a powerful tool for characterizing symmetry-protected topological states and their interaction effect.
\end{abstract}

\maketitle


{\it Introduction.--}
The geometry of quantum states is characterized by the quantum geometric tensor (QGT) whose symmetric real and antisymmetric imaginary parts
correspond to the quantum metric (QM) and Berry curvature (BC), respectively.
The integral of local geometric quantities such as QM and BC often gives information about the global topology~\cite{smooth,Chern1,Chern2,Chern3,Chern4,Chern5,Chern6,Chern7,Chern8,Inequality3,GEuler}.
A well-known example is the first Chern number that is given by the integral of the BC over a closed two-dimensional (2D) space.
The first Chern number governs the topological properties of 2D insulators in the absence of symmetry constraints~\cite{smooth,Chern1,Chern2,Chern3,Chern4,Chern5,Chern6,Chern7,Chern8},
which clearly demonstrates the intimate relationship between the geometry and topology of quantum states.

Not only the BC, but the QM also carries information about global topology. 
For instance, the quantum volume, the volume of the parameter space measured by the QM, has a lower bound determined by the first Chern number~\cite{Inequality, Inequality1, Inequality2, Inequality3, chernAndVolume}. 
However, when the system is under certain symmetry constraints, its global topology is not necessarily characterized by the first Chern number as shown in various symmetry-protected topological states, while the definition of the quantum volumes remains the same~\cite{Sym1,Sym2,Sym3,Sym4}. 
Therefore, it remains a question whether symmetry-protected topology can give a bound of the quantum volume or even be related to it. 
In particular, when symmetry constraints force the BC to be strictly zero, revealing the relationship of the quantum volume to global topology is an intriguing open question.

In fact, the relation between the quantum volume and the Chern number arises from the fundamental local inequality between the QM and BC.
In particular, when the local inequality saturates, the corresponding Chern band is expected to host fractional Chern insulators when interaction effect is included~\cite{Frac1,Frac2,Frac3,Frac4}. Thus, extending the fundamental local inequality between the QM and BC to systems with vanishing BC is an important step towards the complete understanding of many-body instabilities in interacting symmetry-protected topological bands.

In this Letter, we establish the relationship between the quantum metric and Euler band topology in 2D systems with space-time inversion $I_{ST}$ symmetry.
When the antiunitary $I_{ST}$ symmetry satisfying $I_{ST}^2=1$ exists, there is a basis in which the Bloch Hamiltonian and the relevant wave functions become real. Thus, the BC vanishes at every momentum and the first Chern number is always zero.
However, interestingly, isolated two bands in $I_{ST}$ symmetric systems carry another integer $\mathbb{Z}$ topological invariant, called the Euler number $e_2$~\cite{Euler,Euler1,Euler2,EulerAndChern,eulerDefinition}.
Here, we derive the fundamental local inequality between the QM and Euler curvature. Based on this, we establish the relation between the quantum volume and the Euler number. Moreover, from the saturation conditon of the local inequality, we derive the ideal conditon for topological Euler bands, and demonstrate the band geometric criterion on the correlation effect in interacting Euler bands.
Considering the recent discovery of the Euler band topology in the nearly flat bands of twisted bilayer graphene (TBG) at magic angles, 
our theory will shed light on the fascinating correlated topological properties of magic angle TBGs~\cite{Fascinating,superconductivity,oneThirdFilling,TBGExperiment,TBGBKT,TBGCalculation}.

{\it Quantum geometry of Chern bands.---}
Generally, the non-Abelian QGT is given by
\begin{align}\label{eq:QGT}
	\begin{split}
		Q_{\mu\nu}^{ij}(\textbf{k})=\bra{\partial_\mu u_i(\textbf{k})}(1-P(\textbf{k}))\ket{\partial_\nu u_j(\textbf{k})},\\
	\end{split}
\end{align}
where $\mu,\nu=x,y,z$ denote spatial coordinates, $\partial_{\mu}=\partial/\partial k_\mu$, and $P(\textbf{k})=\sum_{i=1}^N \ket{u_i(\textbf{k})}\bra{u_i(\textbf{k})}$ indicates the projection operator to the space spanned by the states \{$\ket{u_{1,...,N}(\textbf{k})}$\}~\cite{nonAbelianQGT,nonAbelian1,nonAbelian2,nonAbelian3,nonAbelian4,geometry,Berry}. 
The quantum metric $g_{\mu\nu}(\textbf{k})\equiv\frac{1}{2}\Tr[Q_{\mu\nu}+Q_{\nu\mu}]$
and the Berry curvature $\Omega_{\mu\nu}(\textbf{k})\equiv i\Tr[ Q_{\mu\nu}-Q_{\nu\mu}]$ with the trace $\Tr$ over band indices 
are generally invariant under $U(N)$ gauge transformation for complex wave functions.

The BC $\Omega^n(\textbf{k})$ and QM $g^n_{ij}(\textbf{k})$ of a band with index $n$ satisfy
the following inequality~\cite{chernAndVolume}
\begin{align}\label{eq:Chern}
	\sqrt{\det(g^n_{\mu\nu}(\textbf{k}))}\geq\frac{1}{2}\left|\Omega^n(\textbf{k})\right|,
\end{align}
whose physical significance can be understand in the following context.
First, when both sides of the inequality are integrated over 2D Brillouin zone (BZ),
we obtain the relation $vol_g\geq \pi |C|$ where $C$ is the Chern number and $vol_g$ is the quantum volume defined as $vol_g\equiv\int d^2\textbf{k} \sqrt{\det(g^n_{\mu\nu}(\textbf{k}))}$ which is the volume of the parameter space computed using QM as the metric. $vol_g$ is an excellent measure of the Chern band topology in many systems including the Landau levels and two-band Hamiltonians where the equality $vol_g=\pi |C|$ holds as well as the flat band systems where $vol_g\approx\pi |C|$ \cite{chernAndVolume}.

Second, the saturation of the inequality in Eq.~(\ref{eq:Chern}) provides an important criterion to achieve fractional topological phases in interacting Chern bands \cite{FCI,FCI2,FCI3}.
The saturation of the inequality with nonzero BC means that the QGT $Q^n_{\mu\nu}(\textbf{k})=g_{\mu\nu}^n(\textbf{k})-i\frac{1}{2}\epsilon_{\mu\nu}\Omega^n(\textbf{k})$ has a null vector, which in turn gives the following relation
$g_{\mu\nu}^n(\textbf{k})=\frac{1}{2}\Omega^n(\textbf{k})\omega_{\mu\nu}(\textbf{k})$
where $\omega_{\mu\nu}(\textbf{k})$ is a $\textbf{k}$-dependent symmetric matrix with unit determinant (see SM) \cite{Landau}. To mimic the lowest Landau level (LLL) under uniform magnetic field, we further assume $\textbf{k}$-independence of $\omega_{\mu\nu}(\textbf{k})$ leading to the following {\it ideal condition} for Chern bands     
\begin{align}
	\label{eq:idealChern}
	g_{\mu\nu}^n(\textbf{k})&=\frac{1}{2}\Omega^n(\textbf{k})\omega_{\mu\nu},
\end{align}
where $\omega_{\mu\nu}$ is a constant symmetric matrix with unit determinant~\cite{magic,Landau}. If Eq.~(\ref{eq:idealChern}) is satisfied, the equality in Eq.~(\ref{eq:Chern}) also holds. 
According to ~\cite{Landau, Complex}, the Bloch wave function $\ket{u(\textbf{k})}$ of an ideal Chern band can be decomposed, like LLL wave functions, as 
\begin{align}
	\label{eq:holomorphic}
	\ket{u(\textbf{k})}=\frac{1}{N_{\textbf{k}}}\ket{\tilde{u}(k)},
\end{align}
where $\ket{\tilde{u}(k)}$ is a holomorphic function of a complex number $k\equiv\lambda_x\textbf{k}_x+\lambda_y\textbf{k}_x$ and $N_{\textbf{k}}$ is the normalization factor.
The complex numbers $\lambda_{x,y}$ satisfy $\omega_{\mu\nu}=\lambda_x^*\lambda_y+\lambda_x\lambda_y^*$ and $i\epsilon_{\mu\nu}=\lambda_\mu^*\lambda_\nu-\lambda_\mu\lambda_\nu^*$ where $\epsilon_{\mu\nu}$ is a fully antisymmetric tensor.

Moreover, when the BC of an ideal Chern band is constant, the corresponding projected density operator $\bar{\rho}_n(\textbf{k})=P_ne^{i\textbf{k}\cdot\textbf{r}}P_n$ where $P_n=\int \ket{u_n(\textbf{k})}\bra{u_n(\textbf{k})}d\textbf{k}$ satisfy so-called the Girvin-MacDonald-Platzman (GMP) algebra, 
\begin{align}
	\begin{split}
		&[\bar{\rho}_n(\textbf{k}),\bar{\rho}_n(\textbf{q})]\\
		&=2ie^{\textbf{k}_\mu g_{\mu\nu}^n\textbf{q}_\nu}\sin\left(\left(\textbf{k}_x\textbf{q}_y-\textbf{k}_y\textbf{q}_x\right)\Omega_{xy}^n\right)\bar{\rho}_n(\textbf{k}+\textbf{q}),
	\end{split}
\end{align}
as in LLL \cite{FCI0,FCI1}, which indicates the stability of the many-body ground state with fractional topology~\cite{FCI,FCI0,FCI1,FCI2,FCI3}. 
Interestingly, recent numerical studies have shown that what is essential to achieve fractional Chern band is not the constant BC but the ideal condition in Eq.~(\ref{eq:idealChern})~\cite{Landau,FCI}.

{\it Nonabelian quantum geometry of real two bands.--}
In two-dimensions, $I_{ST}$ symmetry appears in the form of $I_{ST}=PT$ with time-reversal $T$ and inversion $P$ symmetries in spinless fermion systems, or $I_{ST}=C_{2z}T$ with two-fold rotation $C_{2z}$ symmetry about the $z$-axis in both spinless and spinful fermion systems.
As the antiunitary $I_{ST}$ symmetry is local in momentum space and satisfies  $I^2_{ST}=1$, it can be represented by $I_{ST}=K$ with the complex conjugation operator $K$.
Then, the $I_{ST}$ symmetry of the wave function $\ket{u_i(\textbf{k})}$ ($i$ is a band index) imposes the reality condition 
$I_{ST}\ket{u_i(\textbf{k})}=\ket{u_i(\textbf{k})}^*=\ket{u_i(\textbf{k})}$, which forces the Berry curvature to vanish at every momentum $\textbf{k}$. Thus, the Chern number of $I_{ST}$ symmetric systems is always zero. 
However, two real bands can have nontrivial band topology characterized by the integer Euler invariant as explained below.

For two real bands $\ket{u_{1,2}(\textbf{k})}$, the invariance of the non-Abelian QGT under $O(2)$ gauge transformation leaves
$g_{\mu\nu}(\textbf{k})=Q_{\mu\nu}^{11}(\textbf{k})+Q_{\mu\nu}^{22}(\textbf{k})$ as the only gauge invariant combination.
On the other hand, under $SO(2)$ gauge transformation that preserves the orientation of two real bands, one can find another gauge invariant linear combination $Q_{\mu\nu}^{12}(\textbf{k})-Q_{\mu\nu}^{21}(\textbf{k})$,
which gives the off-digonal Berry curvature $F_{12}(\textbf{k})\equiv Q_{xy}^{12}(\textbf{k})-Q_{yx}^{21}(\textbf{k})=\nabla\times\bra{u_1(\textbf{k})}\nabla\ket{u_2(\textbf{k})}$.
When the orientation of two real bands is fixed, the integral of $F_{12}(\textbf{k})$ becomes the Euler invariant 
\begin{align}\label{eq:Euler}
e_2=\frac{1}{2\pi}\int_{BZ}{d^2\textbf{k}}F_{12}(\textbf{k}),
\end{align}
which classifies the topology of orientable real two bands
~\cite{Euler,Euler1,Euler2,EulerAndChern,eulerDefinition}.
For real bands $\ket{u_{1,2}(\textbf{k})}$,
one can find a Chern basis $\ket{u_{\pm}(\textbf{k})}=\frac{1}{\sqrt{2}}\left(\ket{u_1(\textbf{k})}\pm i\ket{u_2(\textbf{k})}\right)$ satisfying that
when $\ket{u_{1,2}(\textbf{k})}$ have the Euler number $e_2$, $\ket{u_{\pm}(\textbf{k})}$ have the Chern numbers $\pm e_2$, respectively.

{\it Quantum volume and topology of Euler bands.---}
The local geometric quantities $g_{\mu\nu}(\textbf{k})$ and $F_{12}(\textbf{k})$ of two real bands $\ket{u_{1,2}(\textbf{k})}$ satisfy the inequality
\begin{align}\label{eq:Inequality}
  \sqrt{\det(g_{\mu\nu}(\textbf{k}))}\geq \left|F_{12}\left(\textbf{k}\right)\right|,
\end{align}
as shown in the Supplemental Materials (SM).
By integrating both sides over 2D BZ, we obtain
\begin{align}
vol_g\geq \int d^2\textbf{k}\left|F_{12}\left(\textbf{k}\right)\right|\geq \left|{\int d^2\textbf{k}F_{12}\left(\textbf{k}\right)}\right| \geq 2\pi|e_2|.
\end{align}
As a direct consequence of this inequality, if $vol_g<2\pi$, then $|e_2|=0$, i.e., isolated two real bands are topologically trivial. 
In systems with $vol_g\geq2\pi$, the inequality does not give definite information about topology. However, as shown below with several examples, $vol_g/2\pi$ often gives an excellent estimate of topology and approaches the Euler number from above in a proper limit.

{\it Three-band models.--}
Let us illustrate the relationship between the Euler topology and quantum volume by constructing minimal model Hamiltonians. 
Since two real bands with nonzero Euler invariant have fragile Wannier obstruction \cite{Fragile1,Fragile2,Fragile3,Fragile4,Fragile5,Fragile6}, the minimal lattice model for an Euler insulator should have at least three bands with two Euler bands (bands 1 and 2) decoupled from the third band (band 3).

In general, such three-band models have two special properties. First, there must be a point in the BZ where $\det(g_{\mu\nu}(\textbf{k}))=\left|F_{12}(\textbf{k})\right|=0$ for two Euler bands as proved in SM. As a consequence, neither $g_{\mu\nu}(\textbf{k})$ nor $F_{12}(\textbf{k})$ can be nonzero while remaining uniform in the BZ. 
Second, the equality $\sqrt{\det(g_{\mu\nu}(\textbf{k}))}=|F_{12}(\textbf{k})|$ always holds as shown in SM.  A simple consequence is that, if $F_{12}$ has the same sign over the entire BZ, $vol_g = 2\pi|e_2|$. In this case, the quantum volume directly measures $e_2$. 

A three-band model possessing two bands with $e_2\neq0$  can be constructed by using a two-band Chern insulator model
\begin{align}\label{eq:2BandChern}
	H_{Chern}(\textbf{k})=a(\textbf{k})\sigma_x+b(\textbf{k})\sigma_y+c(\textbf{k})\sigma_z,
\end{align}
where $a(\textbf{k})$, $b(\textbf{k})$, $c(\textbf{k})$ are real functions and $\sigma_{x,y,z}$ are Pauli matrices. 
The corresponding real three-band model is given by
\begin{align}\label{eq:3bandEuler}
	H_{Euler}(\textbf{k})=
	\begin{pmatrix}
		a(\textbf{k})^2 & a(\textbf{k})b(\textbf{k}) & a(\textbf{k})c(\textbf{k}) \\
		a(\textbf{k})b(\textbf{k}) & b(\textbf{k})^2 & b(\textbf{k})c(\textbf{k}) \\
		a(\textbf{k})c(\textbf{k}) & b(\textbf{k})c(\textbf{k}) & c(\textbf{k})^2 \\
	\end{pmatrix},
\end{align}
which has two degenerate flat bands at zero energy and a dispersive band with energy $a(\textbf{k})^2+b(\textbf{k})^2+c(\textbf{k})^2$. The BC $\Omega(\textbf{k})$ of Eq.~(\ref{eq:2BandChern}) and the off-diagonal BC $F_{12}(\textbf{k})$ of the two flat bands of Eq.~(\ref{eq:3bandEuler}) satisfy
\begin{align}
	|F_{12}(\textbf{k})|=2|\Omega(\textbf{k})|,
\end{align}
as proved in SM. 
Let us note that regardless of the band dispersion of the Chern insulator, the resulting two Euler bands are perfectly flat. 

\begin{figure}
	\includegraphics[width=0.47\textwidth]{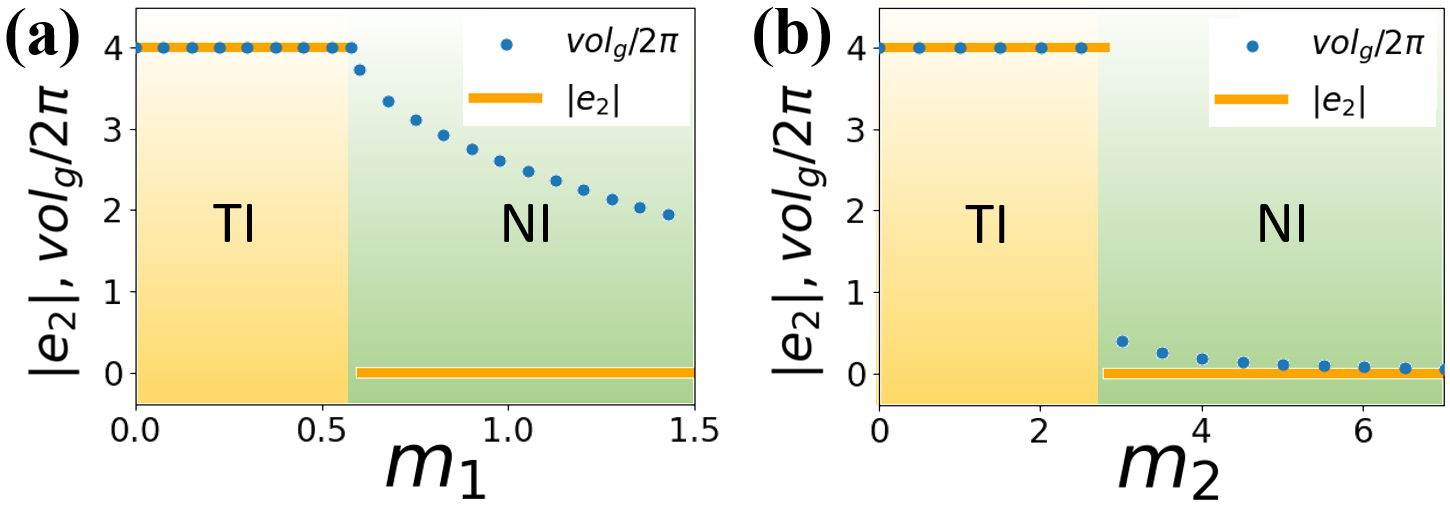}
  \vspace{-0.3cm}
	\caption{(a, b) The change of the quantum volume and Euler invariant of the lower two flat bands for three-band models where a direct transition between a topological insulator (TI) with $e_2\neq0$ and a normal insulator (NI) with $e_2=0$ occurs.
	}
	\label{fig:threeBand}
\end{figure}

For example, let us consider a variant of the square lattice Chern insulator model introduced in ~\cite{Model} with
$a(\textbf{k})=(2-\sqrt{2})t\sin(k_x)\sin(k_y)-m_2$, $b(\textbf{k})=\sqrt{2}t(\cos(k_y)+\cos(k_x))-m_1$,
$c(\textbf{k})=\sqrt{2}t(\cos(k_y)-\cos(k_x))$.
When $m_2=0$, the Chern number of the lower band is $-2$ (0) when $\left|m_1\right|<2\sqrt{2}\left|t\right|$ ($\left|m_1\right|>2\sqrt{2}\left|t\right|$). Accordingly, two degenerate flat bands of the corresponding three-band model have $e_2=4$ ($e_2=0$) when $\left|m_1\right|<2\sqrt{2}\left|t\right|$ ($\left|m_1\right|>2\sqrt{2}\left|t\right|$). 
The change of $vol_g/2\pi$ and $e_2$ as a function of $m_1$ is plotted in Fig.~\ref{fig:threeBand}(a).
Fig.~\ref{fig:threeBand}(b) is a similar plot when $m_2$ is varied with $m_1=0$. In both cases, $vol_g/2\pi$ is an excellent approximation of $e_2$ when $e_2\neq0$. But depending on which parameter is changed, the quantum volume can change either continuously or discontinuously. 

We note that when the degeneracy of the two flat Euler bands is lifted, a topological phase transition changing $e_2$ is mediated by an intermediate semimetal phase~\cite{Euler}. 
Interestingly, although $e_2$ is not well defined in the gapless region, when gap closing points have linear dispersion, the quantum volume is finite and changes smoothly even in the gapless region, as shown in SM.

{\it Multi-band models.--}
Here, we propose a general way to construct a $2N$-band model with two Euler bands decoupled from other bands
by superposing two $N$-band Hamiltonians with an isolated Chern band ($N$ is an integer).
One advantage of this construction is that the well-established inequality in Eq.~(\ref{eq:Chern}) and its saturation condition
for a Chern band~\cite{chernAndVolume}
naturally extend to similar relations for Euler bands.

Explicitly, let us superpose an $N$-band Hamiltonian $H_{Chern}(\textbf{k})$ having an isolated Chern band with its complex conjugate as
\begin{align}\label{eq:Map}
	H_{Euler}(\textbf{k})= 
	\begin{pmatrix}
		H_{Chern}(\textbf{k}) & 0\\
		0 & H_{Chern}^*(\textbf{k})\\
	\end{pmatrix}. 
\end{align}
Then, each band of $H_{Euler}$ is doubly degenerate, and 
the Chern band of $H_{Chern}(\textbf{k})$ turns into degenerate Euler bands of $H_{Euler}$.
After a unitary transformation described in SM,
$H_{Euler}(\textbf{k})$ can become real and commute with a matrix $\tau_y$ as
\begin{align}\label{eq:realMap}
	\tilde{H}_{Euler}(\textbf{k})= \Re[H_{Chern}(\textbf{k})]\otimes \tau_0- \Im[H_{Chern}(\textbf{k})]\otimes i\tau_y,
\end{align}
where $\tau_{x,y,z}$ are Pauli matrices connecting two $N$-band Hamiltonians and $\tau_0$ is the relevant $2\times2$ identity matrix.

The BC $\Omega(\textbf{k})$ and QM $g^{Chern}_{ij}(\textbf{k})$ of the Chern band in $H_{Chern}$
and the Euler curvature $F_{12}(\textbf{k})$ and QM $g^{Euler}_{ij}(\textbf{k})$ of the Euler bands in $\tilde{H}_{Euler}(\textbf{k})$ satisfy the following relation
\begin{align}\label{eq:correpondence}
	\begin{split}
		|F_{12}(\textbf{k})|&=|\Omega(\textbf{k})|,\\
		\sqrt{\det\left(g^{Euler}_{ij}(\textbf{k})\right)}&=2\sqrt{\det\left(g^{Chern}_{ij}(\textbf{k})\right)},\\
	\end{split}
\end{align}
as proved in SM. 
This clearly demonstrates that the band topology of the mapped Euler bands can be well approximated by the quantum volume in the same manner as the Chern band case. 

For example, let us construct a four-band model which can be mapped to two superposed two-band Chern insulators in certain limits.
In such limits, as the inequality in Eq.~(\ref{eq:Chern}) becomes the equality for two-band Chern insulators, a similar equality should hold for Euler bands.
Explicitly, we consider the Hamiltonian $H^{a}_{4}(\textbf{k})$ with components
$H^{a}_{4}(\textbf{k})=a(\textbf{k})\sigma_x+b(\textbf{k})\tau_y\otimes\sigma_y+c(\textbf{k})\sigma_z+m_3\tau_x\otimes\sigma_x$ where $\sigma_{x,y,z,}$, $\tau_{x,y,z}$ are Pauli matrices and $a(\textbf{k}),b(\textbf{k}),c(\textbf{k})$ are the same as those in the previous three-band model with $t=1$ and $m_1=m_2=0$. 
The band dispersion at $m_3=0$ is shown in Fig.~\ref{fig:fourBand}(a).
When $m_3=0$, the Hamiltonian commutes with $\tau_y$, and thus can be written as in Eq.~(\ref{eq:realMap}). 
As shown in Fig.~\ref{fig:fourBand}(c),  $vol_g/2\pi$ and $|e_2|$ of lower degenerate bands coincide only at $m_3=0$. The same also holds for upper degenerate bands.

\begin{figure}
	\includegraphics[width=0.45\textwidth]{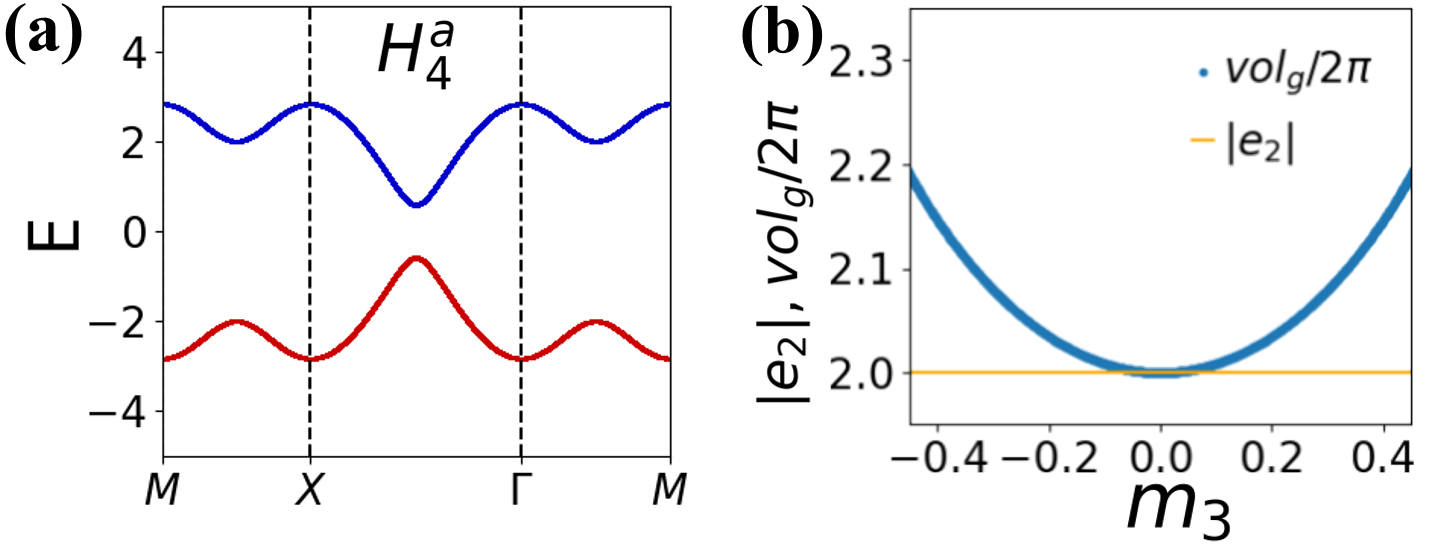}
 \vspace{-0.3cm}
	\caption{(a) The band structure of the four-band model $H^{a}_{4}(\textbf{k})$ when $m_3=0$. Both the red and blue bands are doubly degenerate when $m_3=0$ while the degeneracy is lifted when $m_3\neq 0$. 
	(b) The quantum volume and Euler invariant of the red band in (a). 
}
	\label{fig:fourBand}
\end{figure}

When the real Hamiltonian hosting Euler bands has additional chiral $S$ symmetry satisfying $S^2=1$ and $[S,I_{ST}]=0$, two Euler bands can apppear as isolated zero energy flat bands when $\Tr[S]=\pm2$. Especially, the zero energy states of a $(2N+2)$-band real Hamiltonian with chiral symmetry can be considered as those of a related $(N+2)$-band Hamiltonian sharing the same Euler band topology (see SM).
This means that a 4-band chiral symmetric Hamiltonian with zero energy flat Euler bands, satisfying $\sqrt{\det(g_{\mu\nu}(\textbf{k}))}=|F_{12}(\textbf{k})|$ as in 3-band models, can also be constructed.

On the other hand, if $\{S,I_{ST}\}=0$ is satisfied, isolated two bands cannot have $e_2\neq0$
in periodic real Hamiltonians as shown in SM.
One exception is when Euler bands appear as effective low energy bands of quasi-periodic systems
as in TBG discussed below.


{\it Ideal Euler bands and correlation effect.--} 
Here, we address the question of possible fractional topological phases in interacting Euler bands from the band geometric point of view.
First, we define {\it ideal Euler bands} as two real bands $\ket{u_{1,2}(\textbf{k})}$ whose QGT satisfies
\begin{align}
	\label{eq:idealEuler}
g_{\mu\nu}(\textbf{k})&=F_{12}(\textbf{k})\omega_{\mu\nu},
\end{align}
which is equivalent to the ideal condition for Chern bands in Eq.~(\ref{eq:idealChern}).
Ideal Euler bands always satisfy the equality of Eq.~(\ref{eq:Inequality}).
We note that the corresponding Chern basis $\ket{u_{\pm}(\textbf{k})}=\frac{1}{\sqrt{2}}\left(\ket{u_1(\textbf{k})}\pm i\ket{u_2(\textbf{k})}\right)$ satisfy the ideal condition in Eq.~(\ref{eq:idealChern}).

Since Eq.~(\ref{eq:idealEuler}) is equivalent to the condition that the QGT $G_{\mu\nu}(\textbf{k})\equiv g_{\mu\nu}(\textbf{k})+i\epsilon_{\mu\nu}F_{12}(\textbf{k})$ with $F_{12}(\textbf{k})\neq0$ has a constant null vector, the wave function for ideal Euler bands and the corresponding Chern basis
can always be written as in Eq.~(\ref{eq:holomorphic}), analogous to LLL (see SM). 
This indicates the potential that
fractional topological insulators may appear in partially filled interacting ideal Euler bands~\cite{Landau}.
Moreover, when $F_{12}(\textbf{k})$ of ideal Euler bands is constant in momentum space, the projected density operators $\bar{\rho}_{\alpha\beta}(\textbf{k})=P_\alpha e^{i\textbf{k}\cdot\textbf{r}}P_\beta$ where $P_\alpha=\int \ket{u_\alpha(\textbf{k})}\bra{u_\alpha(\textbf{k})}d\textbf{k}$ ($\alpha=\pm$) for the Chern basis $\ket{u_{\pm}(\textbf{k})}$ also satisfy the algebraic relation similar to the GMP algebra when $\alpha=\beta$, which further supports the possible fractional topological phases. 
However, as $\bar{\rho}_{\alpha\beta}(\textbf{k})$ with $\alpha\neq\beta$ do not satisfy closed algebraic relations, ideal Euler bands is different from two copies of ideal Chern bands, thus one may expect distinct many-body ground states in interacting Euler bands (see SM).

Unfortunately, both ideal Chern and ideal Euler bands cannot be realized in periodic lattice Hamiltonians in which the atomic positions are given by a linear combination of primitive lattice vectors with rational coefficients, which includes most of known lattice systems.
This happens because the decomposition in Eq.~(\ref{eq:holomorphic}) is not compatible with nonzero Chern or Euler number, as shown in SM.
However, ideal bands can be realized in continuum models, as an effective periodic low energy model of quasi-periodic systems such as TBG. 

In general, one can construct a continuum model hosting ideal Euler bands as follows.
For given ideal Euler bands $\ket{u_{1,2}(\textbf{k})}$, the Chern basis $\ket{u_{\pm}(\textbf{k})}$ can be written as
\begin{align}
	\label{eq:ChernBasis}
	\ket{u_{+}(\textbf{k})}=\left(\ket{u_{-}(\textbf{k})}\right)^*=\frac{1}{N_{\textbf{k}}}\ket{\tilde{u}(k)},
\end{align}
where $N_{\textbf{k}}$ is the normalizaton factor. The holomorphic function $\ket{\tilde{u}(k)}$ satisfies
i) $(\bra{\tilde{u}(k)})^*\ket{\tilde{u}(k)}=0$ because of the normalization of $\ket{u_{1,2}(\textbf{k})}$,
and ii) $\forall k,~\bra{\tilde{u}(k)}\ket{\tilde{u}(k)}\neq 0$ because of the normalization condition in Eq.~(\ref{eq:ChernBasis}).
Using $\ket{u_{+}(\textbf{k})}$ and $\ket{\tilde{u}(k)}$, the Euler curvature and quantum metric of the Euler bands can be calculated as
\begin{align}
	\label{eq:HolomorphicEuler}
	&F_{12}(\textbf{k})=i(\lambda_y^*\lambda_x-\lambda_x^*\lambda_y)\bra{v(\textbf{k})}\ket{v(\textbf{k})}=\bra{v(\textbf{k})}\ket{v(\textbf{k})},
	\nonumber\\
	&g_{\alpha\beta}(\textbf{k})=
	(\lambda_\alpha^*\lambda_\beta+\lambda_\beta^*\lambda_\alpha)\bra{v(\textbf{k})}\ket{v(\textbf{k})}=\omega_{\alpha\beta}\bra{v(\textbf{k})}\ket{v(\textbf{k})},
\end{align}
where
\begin{align}
	\ket{v(\textbf{k})}=\frac{1}{N_{\textbf{k}}}(1-\ket{u_+(\textbf{k})}\bra{u_+(\textbf{k})})\ket{\partial_k\tilde{u}(k)}.
\end{align}
In Eq.~(\ref{eq:HolomorphicEuler}), the ideality of Euler bands manifests.

Now, let us construct a 3-band continuum model hosting ideal Euler bands.
Defining $\ket{\tilde{u}(k)}^{t}=\left(\tilde{u}_1(k), \tilde{u}_2(k), \tilde{u}_3(k)\right)$,
one can choose $\tilde{u}_1(k)=f_1(k)^2-f_2(k)^2$, $\tilde{u}_2(k)=i(f_1(k)^2+f_2(k)^2)$, $\tilde{u}_3(k)=2f_1(k)f_2(k)$
where $f_{1,2}(k)$ are analytic polynomial functions of $k$.
The above choice of $\tilde{u}_1(k), \tilde{u}_2(k), \tilde{u}_3(k)$ satisfies two conditions i) $(\bra{\tilde{u}(k)})^*\ket{\tilde{u}(k)}=0$
and ii) $\bra{\tilde{u}(k)}\ket{\tilde{u}(k)}\neq 0$ for any $k$.
The Hamiltonian $H(\textbf{k})=\ket{z(k)}\bra{z(k)}$ with
\begin{align}
	\ket{z(k)}=
	\begin{pmatrix}
		f_1(k)^*f_2(k)+f_1(k)f_2(k)^*\\
		\frac{1}{i}(f_1(k)^*f_2(k)-f_1(k)f_2(k)^*)\\
		|f_2(k)|^2-|f_1(k)|^2
	\end{pmatrix},
\end{align}
has zero energy degenerate flat ideal Euler bands and the third band with energy $(|f_1(k)|^2+|f_2(k)|^2)^2>0$.  
Choosing $f_1(k)=k$, $f_2(k)=1$, the Euler bands have $e_2=\pm2$.
See SM for more general discussion.

\begin{figure}
	\includegraphics[width=0.47\textwidth]{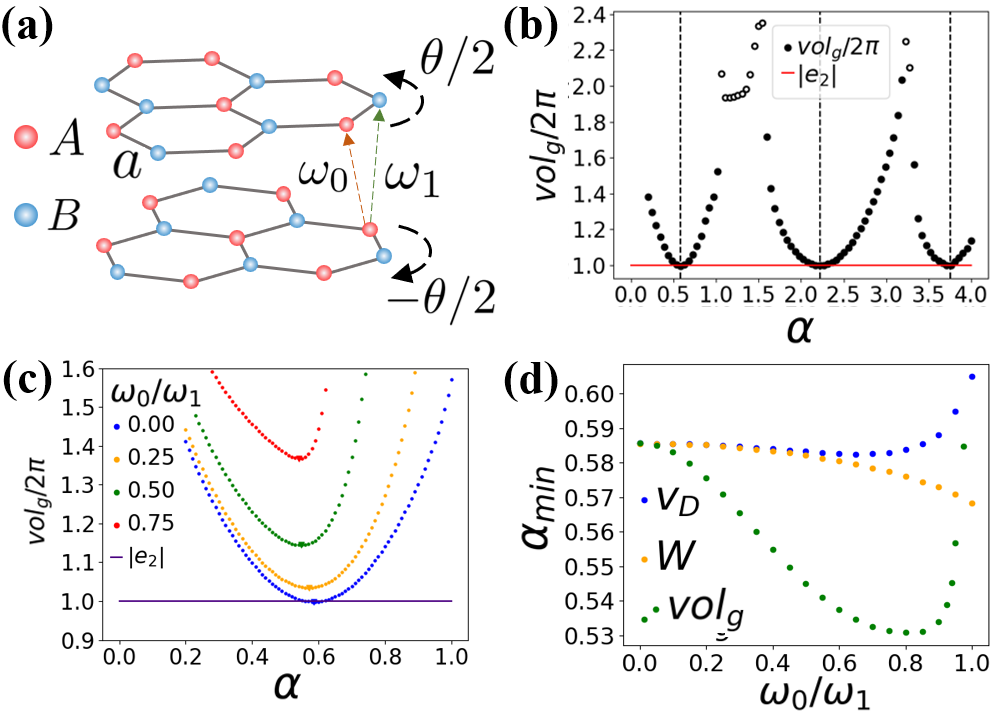}
	\vspace{-0.3cm}
	\caption{(a) The lattice structure of TBG with the twist angle $\theta$. 
		(b) The change of the quantum volume and Euler invariant as a function of $\alpha$ at the chiral limit with $\omega_0/\omega_1=0$. $vol_g/2\pi=|e_2|$ holds only at magic angles. 
		(c) Similar plot for different $\omega_0/\omega_1$. 
		(d) Plot of $\alpha_{\text{min}}$ at which the Dirac velocity $v_D$, the flat band bandwidth $W$, or $vol_g$ becomes minimized at a given $\omega_0/\omega_1$ as a function of $\omega_0/\omega_1$.
	}
	\label{fig:TBG}
\end{figure}

{\it Twisted bilayer graphene (TBG).--}
The nearly flat bands at charge neutrality in TBG with small twist angle $\theta$
are the representative example of Euler bands with $e_2=1$~\cite{Euler},
which can be described by the continuum model $H_{BM}$ proposed by Bistritzer and MacDonald~\cite{Moire}.
The low energy band structure of $H_{BM}$ can be characterized by two dimensionless parameters $\omega_0/\omega_1$ and $\alpha=\omega_1/(v_0k_\theta)$ 
where $\omega_0$ and $\omega_1$ describe the interlayer couplings between AA/BB sites and AB/BA sites, respectively [see Fig.~\ref{fig:TBG}(a)]~\cite{Moire}.
$v_0$ is the Fermi velocity of Dirac points and $k_\theta=8\pi\sin(\theta/2)/3a$ with lattice constant $a$ (see SM for details).

When $\omega_0/\omega_1=0$, TBG has chiral symmetry $S$ that anticommutes with $C_{2z}T$. 
However, the flat bands in TBG can carry nonzero $e_2$ because of the quasi-periodicity of TBG (see SM).
When chiral symmetry exists, $vol_g/2\pi=e_2=1$ holds at each magic angle with flat bands as shown in Fig.~\ref{fig:TBG}(b). 
 
When chiral symmetry is broken, $vol_g/2\pi>e_2$ holds (Fig.~\ref{fig:TBG}(c)). 
Also, the minimum quantum volume, minimum bandwidth, and minimum Dirac velocity appear at different $\theta$'s (Fig~\ref{fig:TBG}(d)), each of which has its own physical significance. 
Considering interaction effect, however, $\theta$ with minimum quantum volume is special because two quantities on both sides in Eq.~(\ref{eq:Inequality}) are closest at this point, thus correlation induced fractional topological phases should be the most favorable around it.

{\it Conclusion and Discussion.--}
We have shown that the quantum volume provides an excellent measure of the Euler band topology in 2D systems with $I_{ST}$ symmetry in which the BC vanishes, and thus the well-known inequality between the quantum volume and BC in Eq.~(\ref{eq:Chern}) is not meaningful. Moreover, the equality of Eq.~(\ref{eq:Inequality}) was shown to be deeply related to the ideal condition for interacting Euler bands, which may host fractional topological insulators. In the case of TBG in chiral limit, because chiral symmetry satisfies $S^2=1,\{S,I_{ST}\}=0$, the ideal Euler bands at magic angle are reduced to two decoupled ideal Chern bands on which most of the recent theoretical studies on correlation effect are focused.
However, as real TBG systems are not chiral symmetric, the Euler bands and their ideal limit without chiral symmetry might be a more appropriate starting point to examine correlation effects.

Moreover, to observe the genuine properties of interacting ideal Euler bands distinct from interacting ideal Chern bands, finding proper material platforms hosting ideal Euler bands is crucial.
Thorough examination of the many-body instability in interacting ideal Euler bands is an important problem which we leave for future study.

\begin{acknowledgments}
S. K and B.J.Y. were supported by the Institute for Basic Science in Korea (Grant No. IBS-R009-D1), 
Samsung  Science and Technology Foundation under Project Number SSTF-BA2002-06,
and the National Research Foundation of Korea (NRF) grant funded by the Korea government (MSIT) (No. 2021R1A2C4002773, and No. NRF-2021R1A5A1032996).
\end{acknowledgments}







	
 

\appendix
\section{Geometrical and topological properties of wavefunction under the space-time symmetry}
Among various symmetry constraints, space-time inversion symmetry, $I_{ST}$, is local in the Brillouin zone, since both space inversion and time inversion flip the Bloch wave vector. Being local, $I_{ST}$ can give strong constraints to Bloch wavefunctions. For example, when $I_{ST}^2=-1$, by Kramer's theorem, all the bands should be degenerate in the whole Brillouin zone. $I_{ST}^2$ can also be $1$ when spin-orbit coupling is absent. Additionally, for two-dimensional systems, $I_{ST}^2=1$ is possible even in the presence of spin-orbit coupling, when $I_{ST}=C_{2z}T$. In this case, we can define a real gauge as
\begin{align}
    I_{ST}\ket{u_n(\textbf{k})}=\ket{u_n(\textbf{k})},
\end{align}
where $\ket{u_n(\textbf{k})}$ is a Bloch wavefunction. It is called the real gauge because there exists a basis that makes $I_{ST}$ a complex conjugate operator, resulting in the real Bloch wavefunctions~\cite{real1,real2}. The $\mathbb{Z}$ invariant classifying the topology of this real Bloch band is the Euler class. Unlike the Chern class, the Euler class is well-defined only for the two bands that are isolated from other bands. In this section, we examine the geometry of these two isolated real bands to look into the relationship between topology and geometry under space-time inversion symmetry $I_{ST}^2=1$. In the process, we define the Euler class and quantum volume and characterize the relationship between them.

\subsection{Geometric quantity}
Every geometric quantity, such as length, volume, curvature, and connection can be constructed from metric. Therefore, defining and analyzing a gauge-invariant distance between wavefunctions is a good starting point for finding geometric quantities. The non-Abelian quantum geometric tensor can be derived from the distance between two nearby states defined as $\left\|\ket{\psi(\textbf{k}+d\textbf{k})}-\ket{\psi(\textbf{k})} \right\|$. Here $\psi(\textbf{k})=\sum_{i=1}^N C_i\ket{u_i(\textbf{k})}$, where $C_i$ is an arbitrary complex number, is a linear combination of a set of Bloch wavefunctions, $\ket{u_1(\textbf{k})}, ... , \ket{u_N(\textbf{k})}$. The matrix element of the QGT can be found as follows
\begin{align}\label{eq:QGT}
    \begin{split}
        Q_{\mu\nu}^{ij}(\textbf{k})=\bra{\partial_\mu u_i(\textbf{k})}(1-P(\textbf{k}))\ket{\partial_\nu u_j(\textbf{k})},\\
    \end{split}
\end{align}
where $\mu,\nu=x,y$, $\partial_{\mu}=\partial/\partial \textbf{k}_\mu$ and $P(\textbf{k})$ stands for the projection operator, defined as $P(\textbf{k})=\sum_{i=1}^N \ket{u_i(\textbf{k})}\bra{u_i(\textbf{k})}$~\cite{nonAbelianQGT}. This tensor is gauge invariant for $U(1)$ gauge for each band. However, to be well-defined for degenerate bands, geometric quantity should also be invariant for gauges that mix bands. It can be found from linear combinations of the elements of the quantum geometric tensor. Under unitary gauge, band indices should be traced out to be gauge invariant. Using a quantum geometric tensor with band indices traced out, the quantum metric and the Berry curvature can be defined as
\begin{align}\label{eq:MetricCurvature}
    \begin{split}
        &g_{\mu\nu}(\textbf{k})=\frac{1}{2}\left(\sum_{i=1}^N Q_{\mu\nu}^{ii}(\textbf{k})+Q_{\nu\mu}^{ii}(\textbf{k})\right)=Re\left[Q^{ii}_{\mu\nu}(\textbf{k})\right],\\
        &\Omega_{\mu\nu}(\textbf{k})=i\left(\sum_{i=1}^N Q_{\mu\nu}^{ii}(\textbf{k})-Q_{\nu\mu}^{ii}(\textbf{k})\right)=-2Im\left[Q^{ii}_{\mu\nu}(\textbf{k})\right].\\
    \end{split}
\end{align}

For real two bands, the most general gauge that mixes two bands is $O(2)$ gauge. Therefore, geometric quantities should be gauge invariant under $O(2)$ gauge. To find gauge invariant linear combinations, we explicitly write down how the elements of the quantum geometric tensor change by $O(2)$ gauge transformation.
$O(2)$ transformation can be generally written as
\begin{align}
    \begin{split}
    &\begin{pmatrix}
    \ket{u_1(\textbf{k})}\\
    \ket{u_2(\textbf{k})}\\
    \end{pmatrix}\rightarrow O(\textbf{k})
    \begin{pmatrix}
    \ket{u_1(\textbf{k})}\\
    \ket{u_2(\textbf{k})}\\
    \end{pmatrix},\\
    \end{split}
\end{align}
where
\begin{align}
    \begin{split}
    O(\textbf{k})=
    \begin{pmatrix}
    \cos\theta(\textbf{k}) & -\sin\theta(\textbf{k})\\
    \sin\theta(\textbf{k}) & \cos\theta(\textbf{k})\\
    \end{pmatrix}.
    \end{split}
\end{align}
 This transformation changes the non-Abelian quantum geometric tensor as
\begin{align}
\begin{split}
    &\begin{pmatrix}
        &Q_{\mu\nu}^{11}(\textbf{k})\\
        &Q_{\mu\nu}^{12}(\textbf{k})\\
        &Q_{\mu\nu}^{21}(\textbf{k})\\
        &Q_{\mu\nu}^{22}(\textbf{k})\\
    \end{pmatrix}\rightarrow T(\textbf{k})
    \begin{pmatrix}
        &Q_{\mu\nu}^{11}(\textbf{k})\\
        &Q_{\mu\nu}^{12}(\textbf{k})\\
        &Q_{\mu\nu}^{21}(\textbf{k})\\
        &Q_{\mu\nu}^{22}(\textbf{k})\\
    \end{pmatrix},\\
\end{split}
\end{align}
where
\begin{align}
\begin{split}
&T(\textbf{k})=\\
&\begin{pmatrix}
    &\cos^2\theta(\textbf{k}) &-\frac{1}{2}\sin2\theta(\textbf{k}) &-\frac{1}{2}\sin2\theta(\textbf{k}) & \sin^2\theta(\textbf{k})\\
    &\pm\frac{1}{2}\sin2\theta(\textbf{k}) &\pm\cos^2\theta(\textbf{k}) &\mp\sin^2\theta(\textbf{k}) & \mp\frac{1}{2}\sin2\theta(\textbf{k})\\
    &\pm\frac{1}{2}\sin2\theta(\textbf{k}) &\mp\sin^2\theta(\textbf{k}) &\pm\cos^2\theta(\textbf{k}) & \mp\frac{1}{2}\sin2\theta(\textbf{k})\\
    &\sin^2\theta(\textbf{k}) &\frac{1}{2}\sin2\theta(\textbf{k}) &\frac{1}{2}\sin2\theta(\textbf{k}) & \cos^2\theta(\textbf{k})\\
\end{pmatrix}.
\end{split}
\end{align}
Under the same gauge transform, the linear combination of the elements is transformed as

\begin{align}
\begin{split}
    &\begin{pmatrix}
        c_{11} & c_{12} & c_{21} & c_{22}
    \end{pmatrix}
    \begin{pmatrix}
        &Q_{\mu\nu}^{11}(\textbf{k})\\
        &Q_{\mu\nu}^{12}(\textbf{k})\\
        &Q_{\mu\nu}^{21}(\textbf{k})\\
        &Q_{\mu\nu}^{22}(\textbf{k})\\
    \end{pmatrix}\\
    &\rightarrow\begin{pmatrix}
        c_{11} & c_{12} & c_{21} & c_{22}
    \end{pmatrix}
    T(\textbf{k})
    \begin{pmatrix}
        &Q_{\mu\nu}^{11}(\textbf{k})\\
        &Q_{\mu\nu}^{12}(\textbf{k})\\
        &Q_{\mu\nu}^{21}(\textbf{k})\\
        &Q_{\mu\nu}^{22}(\textbf{k})\\
    \end{pmatrix}.\\
\end{split}
\end{align}
Therefore, to be gauge invariant, the equation
\begin{align}
\begin{split}
    \begin{pmatrix}
        c_{11}&c_{12}&c_{21}&c_{22}\\
    \end{pmatrix}T(\textbf{k})=
    \begin{pmatrix}
        c_{11}&c_{12}&c_{21}&c_{22}\\
    \end{pmatrix}
\end{split}
\end{align}
should hold.
The only gauge-invariant linear combination is $Q_{\mu\nu}^{11}(\textbf{k})+Q_{\mu\nu}^{22}(\textbf{k})$. This quantity is also invariant for the unitary gauge since it is a quantum geometric tensor. For real bands, there is no imaginary part of this tensor, therefore the trace of the Berry curvature vanishes and the quantum geometric tensor becomes identical to the quantum metric. Typical geometric quantity for two real bands is connected to topology which is introduced in the next section.
\newline
\newline
\newline

\subsection{\label{sec:level2}Topological quantity}
If the gauge restriction is further restrained to $SO(2)$, there is another invariant linear combination, which is $Q_{\mu\nu}^{12}(\textbf{k})-Q_{\mu\nu}^{21}(\textbf{k})$. For $\mu=\nu$, this becomes zero. The remaining term, $Q_{xy}^{12}(\textbf{k})-Q_{yx}^{21}(\textbf{k})=-(Q_{yx}^{12}(\textbf{k})-Q_{xy}^{21}(\textbf{k}))$, is same as the off-diagonal term of the real Berry curvature, $F_{12}(\textbf{k})=\nabla\times\bra{u_1(\textbf{k})}\nabla\ket{u_2(\textbf{k})}$. Note that the real Berry curvature has no factor $i$ since it is the Berry curvature for real bands. For $O(2)$ gauge this quantity has sign ambiguity caused by orientation reversing transformation with $\det[O(\textbf{k})]=-1$. Therefore, the orientation of real bands should be determined to well define $F_{12}(\textbf{k})$. If the orientation of real bands can be determined, the integral of $F_{12}(\textbf{k})$ becomes quantized. This quantization gives the Euler class, which classifies the topology of orientable two real bands~\cite{Euler}.

\begin{align}\label{eq:Euler}
e_2=\frac{1}{2\pi}\int_{BZ}{d^2\textbf{k}}F_{12}(\textbf{k}).
\end{align}
 Since only the sign of the Euler class is ambiguous for $O(2)$ gauge, the absolute value of the Euler class could have a relationship with other physical quantities even when the orientation of the wave functions is not fixed. It is explained in the next section.

\subsection{Relationship between topology and geometry}
The local geometric quantities found so far in two real bands are quantum metric and the off-diagonal term of the real Berry curvature. The inequality that connects these two quantities is
\begin{align}\label{eq:Inequality}
  \sqrt{\det(g_{\mu\nu}(\textbf{k}))}\geq \left|F_{12}\left(\textbf{k}\right)\right|.
\end{align}
 It can be proven by examining the norm of the following wavefunction
\begin{align}
\ket{\phi}=c_x(1-P(\textbf{k}))\ket{\partial_x u_+(\textbf{k})}+c_y(1-P(\textbf{k}))\ket{\partial_y u_+(\textbf{k})},
\end{align}
where $\ket{u_\pm(\textbf{k})}=\left(\ket{u_1(\textbf{k})}\pm i\ket{u_2(\textbf{k})}\right)/\sqrt{2}$ and $P(\textbf{k})=\ket{u_1(\textbf{k})}\bra{u_1(\textbf{k})}+\ket{u_2(\textbf{k})}\bra{u_2(\textbf{k})}$ is a projector to the two real bands. Since $\ket{u_1(\textbf{k})}, \ket{u_2(\textbf{k})}$ are real, one can see that
\begin{align}
    0\leq\bra{\phi}\ket{\phi}=\sum_{i,j=x,y}c_i^*G_{ij}c_j,
\end{align}
 where $G_{ij}=g_{ij}+i\epsilon_{ij}F_{12}$. It becomes more clear by writing down the definition of $F_{12}$ in terms of wavefunctions
\begin{align}
F_{12}\left(\textbf{k}\right)=\bra{\partial_{x}u_1(\textbf{k})}\ket{\partial_{y}u_2(\textbf{k})}-\bra{\partial_{y}u_1(\textbf{k})}\ket{\partial_{x}u_2(\textbf{k})}.
\end{align}
By the inequality, $G_{ij}$ is positive semi-definite. As a consequence the determinant is semi-positive.
\begin{align}
\label{eq:startOfIdeal}
0\leq \det\left(G_{ij}\right)=\det(g_{\mu\nu}(\textbf{k}))-\left|F_{12}\right|^2.
\end{align}
This proves the inequality. For the inequality to be saturated, the determinant should be zero. Note that it is equivalent with $(1-P(\textbf{k}))\ket{\partial_xu_+(\textbf{k})}$ and $(1-P(\textbf{k}))\ket{\partial_yu_+(\textbf{k})}$ being parallel. For the three-band model, because $(1-P(\textbf{k}))$ projects to one dimensional subspace, $(1-P(\textbf{k}))\ket{\partial_x u_+(\textbf{k})}$ and $(1-P(\textbf{k}))\ket{\partial_y u_+(\textbf{k})}$ are always parallel. Due to this, inequality is always saturated. Three-band Euler insulators have properties other than the saturation of the inequality which is introduced in the next section.

\section{Typical properties for three band Euler insulator}

\subsection{Existence of $\det(g_{\mu\nu}(\textbf{k}))=0$ in real three-band model}\label{appendix:zero}
It is known that even though the dimensions of tangent space of 2-torus and 2-sphere are both two, there is no immersion from the Brillouin zone to the sphere~\cite{Complex} i.e. for a smooth function from 2-torus to 2-sphere, there exists a point in torus which the result of the derivation in various directions are linearly dependent on each other.

Here, we show how to use this to prove that there exists a $\textbf{k}$ for which the determinant of the quantum metric of the two bands isolated from the third band in three-band Hamiltonian becomes zero, $\det(g_{\mu\nu}(\textbf{k}))=0$ in the Brillouin zone. We begin with representing the quantum geometric tensor of isolated two bands by the decoupled band. From Eq.~(\ref{eq:QGT}), 
\begin{align}
    Q_{\mu\nu}^{ij}(\textbf{k})=\bra{\partial_{\mu} u_i}\ket{u_3}\bra{u_3}\ket{\partial_\nu u_j},
\end{align}
where $i,j=1,2$, $\ket{u_1(\textbf{k})}, \ket{u_2(\textbf{k})}$ are the isolated two bands and $\ket{u_3(\textbf{k})}$ is a decoupled band. Because $\bra{\partial_{\mu} u_i}\ket{u_j}=\partial_{\mu}\left(\bra{u_i}\ket{u_j}\right)-\bra{u_i}\ket{\partial_{\mu} u_j}=-\bra{u_i}\ket{\partial_{\mu} u_j}$, it can be rewritten as 
\begin{align}
    Q_{\mu\nu}^{ij}(\textbf{k})=\bra{\partial_\nu u_3}\ket{u_j}\bra{u_i}\ket{\partial_{\mu} u_3}.
\end{align}
For real bands, $\bra{\partial_{\mu} u_i}\ket{u_i}=0$ and $\bra{\partial_{\mu} u_i}\ket{\partial_\nu u_j}=\bra{\partial_\nu u_j}\ket{\partial_{\mu} u_i}$. Therefore, by Eq.~(\ref{eq:MetricCurvature}) quantum metric becomes
\begin{align}
    g_{\mu\nu}(\textbf{k})=\bra{\partial_{\mu} u_3}\ket{\partial_\nu u_3}.
\end{align}
$\ket{\partial_x u_3(\textbf{k})}$, $\ket{\partial_y u_3(\textbf{k})}$ are linearly independent if and only if $\det(g_{\mu\nu}(\textbf{k}))\neq0$. Therefore, when $\det(g_{\mu\nu}(\textbf{k}))\neq0$ for the whole Brillouin zone, $\ket{\partial_x u_3(\textbf{k})}$, $\ket{\partial_y u_3(\textbf{k})}$ spans the rank 2 vector space regardless of $\textbf{k}$. Since the set of the unit vector in three-dimensional space is equivalent to the sphere, if the two derivatives $\ket{\partial_x u_3(\textbf{k})}, \ket{\partial_y u_3(\textbf{k})}$ becomes linearly independent for the whole Brillouin zone, there exists an immersion from the Brillouin zone to the sphere. Because it contradicts the fact that there is no immersion from the Brillouin zone to the sphere, there should be a point in the Brillouin zone where $\det(g_{\mu\nu}(\textbf{k}))=0$.
\newline

\subsection{From the two-band Chern insulator to three-band the Euler insulator}\label{appendix:threeband}

Since adding identity to the Hamiltonian does not change eigenstates, only three variables are needed to describe the topology of the two-band Hamiltonian. 
\begin{align}
\label{eq:ChernTwo}
H_{Chern}(\textbf{k})=a(\textbf{k})\sigma_x+b(\textbf{k})\sigma_y+c(\textbf{k})\sigma_z.
\end{align}
The Chern class for a lower band of this Hamiltonian is the same as the Euler class of degenerate lower two bands of
\begin{align}
\label{eq:EulerThree}
H(\textbf{k})=
\begin{pmatrix}
a(\textbf{k})^2 & a(\textbf{k})b(\textbf{k}) & a(\textbf{k})c(\textbf{k}) \\
a(\textbf{k})b(\textbf{k}) & b(\textbf{k})^2 & b(\textbf{k})c(\textbf{k}) \\
a(\textbf{k})c(\textbf{k}) & b(\textbf{k})c(\textbf{k}) & c(\textbf{k})^2 \\
\end{pmatrix}.
\end{align}
For better parameterization, we use
\begin{align}
\begin{split}
&a(\textbf{k})=r\sin\theta(\textbf{k})\cos\phi(\textbf{k}),\\
&b(\textbf{k})=r\sin\theta(\textbf{k})\sin\phi(\textbf{k}),\\
&c(\textbf{k})=r\cos\theta(\textbf{k}),\\
\end{split}
\end{align}
where $r(\textbf{k})$ is positive. Using this parameterization, the eigenstate for the lower band in the two-band system can be written as $\ket{u_1(\textbf{k})}=(-\sin\frac{\theta}{2},e^{i\phi}\cos\frac{\theta}{2})^T$. The Berry curvature of this wavefunction is given by $\frac{1}{2}(\partial_x\theta\partial_y\phi-\partial_y\theta\partial_x\phi)\sin\theta$. The off-diagonal term of the real Berry curvature of two degenerate lower bands of three-band system can be obtained from real eigenstates $\ket{u_1(\textbf{k})}=(\sin\phi,-\cos\phi,0)^T$, $\ket{u_2(\textbf{k})}=(\cos\theta\cos\phi,\cos\theta\sin\phi,-\sin\theta)^T$. The off-diagonal term of the real Berry curvature of these states is $(\partial_x\theta\partial_y\phi-\partial_y\theta\partial_x\phi)\sin\theta$ which is twice the Berry curvature of the lower Chern band. Since $vol_g=\frac{1}{2}|\Omega|$ for the two-band Chern insulator \cite{chernAndVolume} and $vol_g=F_{12}$ for the three-band real Bloch Hamiltonian, the quantum volume becomes quadruple.

\subsection{Example of three-band Euler insulator}
In this section, we make an example of a three-band Euler insulator using the correspondence between the two-band Chern insulator and the three-band Euler insulator given in the previous section. The variant of the square lattice Chern insulator model in ~\cite{Model} can be represented using Eq. (\ref{eq:ChernTwo}) and
\begin{align}
\begin{split}
&a(\textbf{k})=(2-\sqrt{2})t\sin(k_x)\sin(k_y)-m_2,\\ &b(\textbf{k})=\sqrt{2}t(\cos(k_y)+\cos(k_x))-m_1,\\
&c(\textbf{k})=\sqrt{2}t(\cos(k_y)-\cos(k_x)).   
\end{split}
\end{align}
When $m_2=0$, the Chern number of the lower band is $-2$ (0) when $\left|m_1\right|>2\sqrt{2}\left|t\right|$ ($\left|m_1\right|<2\sqrt{2}\left|t\right|$).

\begin{figure}
	\includegraphics[width=0.47\textwidth]{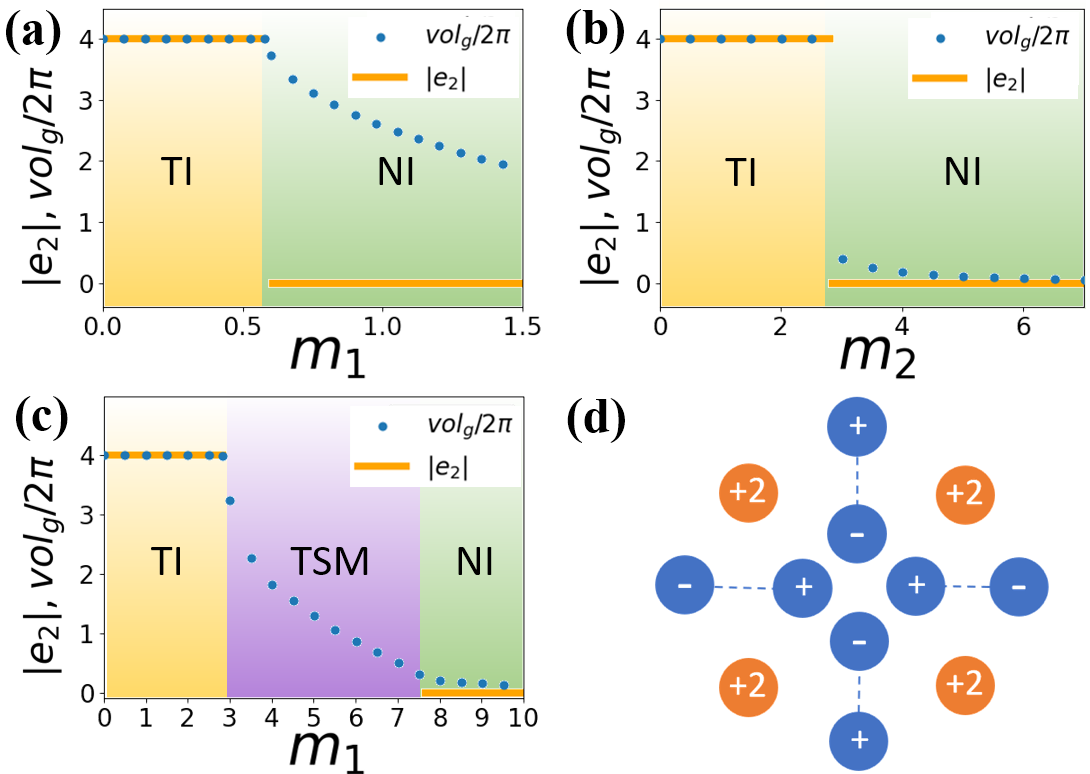}
  \vspace{-0.3cm}
	\caption{(a, b) The change of the quantum volume and Euler invariant of the lower two flat bands for three-band models where a direct insulator-insulator transition occurs.
		(c) Similar plot when the lower two Euler bands are not degenerate where an insulator-semimetal-insulator transition occurs. 
		(d) Distribution of band crossing points and their vorticities in the semimetal phase. 
		The blue (orange) dot is a band touching point between the upper (lower) two bands. 
		As only linear band crossing points with unit vorticity ($\pm$) exist between the upper two bands, the quantum volume of the lower two bands is well-defined.
	}
	\label{fig:threeBand}
\end{figure}

Using the same parameters, a three-band Euler insulator can be constructed using Eq. (\ref{eq:EulerThree}). The two degenerate flat bands of the corresponding three-band model have $e_2=4$ ($e_2=0$) when $\left|m_1\right|<2\sqrt{2}\left|t\right|$ ($\left|m_1\right|>2\sqrt{2}\left|t\right|$). The change of $vol_g/2\pi$ and $e_2$ as a function of $m_1$ is plotted in Fig.~\ref{fig:threeBand}(a).
Fig.~\ref{fig:threeBand}(b) is a similar plot when $m_2$ is varied with $m_1=0$. In both cases, $vol_g/2\pi$ is an excellent approximation of $e_2$ when $e_2\neq0$. However, depending on which parameter is changed, the quantum volume can change either continuously or discontinuously. 

When the degeneracy of the two flat Euler bands is lifted, a topological phase transition changing $e_2$ is mediated by an intermediate Dirac semimetal phase because Dirac points carry quantized $\pi$ Berry phase under $I_{ST}$ symmetry~\cite{Euler}. 
For instance, by adding a constant to the (11) component of Eq.~(\ref{eq:EulerThree}), an insulator-semimetal-insulator transition can occur as shown in Fig.~\ref{fig:threeBand}(c). Interestingly, when band gap closing is linear, although $e_2$ is not well defined in the gapless region, the quantum volume is finite and changes smoothly even in the gapless region. Thus, the quantum volume probes the evolution of the phase transition via an intermediate gapless region.

\section{Condition for finiteness of quantum volume at semimetal}
The Euler number is not well defined when the band gap is closed. The inequality can have meaning in this semimetal phase because when every band gap closing is linear and only one pair of bands meets at each band gap closing point, the quantum volume is well-defined as a finite number. This can be verified by writing the definition of $\det(g_{\mu\nu}(\textbf{k}))$ by the derivative of Hamiltonians.

\begin{widetext}
 \begin{align}
 \det(g_{\mu\nu}(\textbf{k}))=
\frac{1}{2}\sum_{n_1,n_2\in occ,~m_1,m_2\notin occ}{\frac{\left|\bra{u_{m_1}}\partial_xH\ket{u_{n_1}}\bra{u_{m_2}}\partial_yH\ket{u_{n_2}}-\left(n_1,m_1\right)\longleftrightarrow\left(n_2,m_2\right)\right|^2}{\left(E_{n_1}-E_{m_1}\right)^2\left(E_{n_2}-E_{m_2}\right)^2}}.
\end{align}
\end{widetext}

If there is only one pair of $(n,m)$ such that $E_{n}-E_{m}=0$, $n\in occ$, $m\notin occ$ then because there is no term with $\left(n_1,m_1\right)=\left(n_2,m_2\right)$, $\sqrt{\det(g_{\mu\nu}(\textbf{k}))}$ becomes proportional to $\frac{1}{|\textbf{k}-\textbf{k}_0|}$ near band touching point, where $\textbf{k}_0$ is a band gap closing point in the Brillouin zone. This singular point gives a finite contribution for integral in a two-dimensional system, which allows the quantum volume to remain finite.

\section{Role of symmetry constraints}
For the Euler insulator having more than three bands, additional symmetry constraints are important for the relation between the geometry and topology of the Bloch wavefunctions. In this section, we explain how the chiral and mirror symmetry allows the understanding of the relation between topology and geometry in the Euler insulator as the relation in the Euler insulator with a smaller number of bands or as the relation in the Chern insulator.

\subsection{Chiral symmetry $S^2=1$ that commutes with $I_{ST}^2=1$}
For the real gauge, $I_{ST}$ is complex conjugation. Therefore, the chiral symmetry that commutes with $I_{ST}$ should be real. The real chiral symmetry operator with $S^2=1$ can be diagonalized by orthogonal matrices, allowing diagonalization while maintaining the reality of the Bloch Hamiltonian. The block diagonalized chiral symmetry operator will have the form
\begin{align}
    S=
    \begin{pmatrix}
    1_{N\times N} & 0_{N\times M}\\
    0_{M\times N} & -1_{M\times M}\\
    \end{pmatrix},
\end{align}
and correspondingly, the Bloch Hamiltonian takes the form, 
\begin{align}
    H(\textbf{k})=
    \begin{pmatrix}
    0_{N\times N} & A(\textbf{k})\\
    A^{t}(\textbf{k}) & 0_{M\times M}\\
    \end{pmatrix},
\end{align}
where $A(\textbf{k})$ is an $N\times M$ real matrix. For this form of the Bloch Hamiltonian, the eigenvectors can be separated as
\begin{align}
    \begin{pmatrix}
    0_{N\times N} & A(\textbf{k})\\
    A^{t}(\textbf{k}) & 0_{M\times M}\\
    \end{pmatrix}
    \begin{pmatrix}
        \ket{\psi(\textbf{k})}\\
        \ket{\phi(\textbf{k})}\\
    \end{pmatrix}
    =E(\textbf{k})
    \begin{pmatrix}
        \ket{\psi(\textbf{k})}\\
        \ket{\phi(\textbf{k})}\\
    \end{pmatrix},
\end{align}
where $E(\textbf{k})\ket{\psi(\textbf{k})}=A(\textbf{k})\ket{\phi(\textbf{k})}$, $E(\textbf{k})\ket{\phi(\textbf{k})}=A^{t}(\textbf{k})\ket{\psi(\textbf{k})}$. The Bloch eigenstates of the Hamiltonian consist of $\min(M, N)$ chiral symmetric pairs and $|M-N|$ additional flat bands with energy eigenvalue zero. If $M>N+2$, there are at least three bands with zero energy, therefore it is not possible to have two middle bands isolated from other bands. It is the same for the case with $M<N+2$. If $M=N+1$, we have $N$ chiral symmetric pairs and one zero state. Even in this case, we can not have a middle two band that is isolated from other bands, because there can be only an odd number of middle bands that are isolated from other bands. 

When $M=N+2$, at least two orthonormal vectors are the kernel of $A(\textbf{k})$. Therefore the Bloch Hamiltonian has at least two states with zero energy which the eigenvalue equation is given by,
\begin{align}
    \begin{split}
    &\begin{pmatrix}
    0_{N\times N} & A(\textbf{k})\\
    A^{t}(\textbf{k}) & 0_{M\times M}\\
    \end{pmatrix}
    \begin{pmatrix}
        0_{N\times 1}\\
        \ket{\phi_1(\textbf{k})}\\
    \end{pmatrix}\\
    &=\begin{pmatrix}
    0_{N\times N} & A(\textbf{k})\\
    A^{t}(\textbf{k}) & 0_{M\times M}\\
    \end{pmatrix}
    \begin{pmatrix}
        0_{N\times 1}\\
        \ket{\phi_2(\textbf{k})}\\
    \end{pmatrix}
    =0.
    \end{split}
\end{align}
Since the system has chiral symmetry if the middle two bands are isolated from other bands, the middle two bands should be these two bands with zero energy. The quantum metric and Euler curvature for these two bands are given by
\begin{align}
    &g_{ij}(\textbf{k})=Re[\bra{\partial_i \phi_1(\textbf{k})}\ket{\partial_j \phi_1(\textbf{k})}+\bra{\partial_i \phi_2(\textbf{k})}\ket{\partial_j \phi_2(\textbf{k})}]\\
    &F_{12}(\textbf{k})=\bra{\partial_x \phi_1(\textbf{k})}\ket{\partial_y \phi_2(\textbf{k})}-\bra{\partial_y \phi_1(\textbf{k})}\ket{\partial_x \phi_2(\textbf{k})}
\end{align}
, which are the same as the quantum metric and the Euler curvature of $\ket{\phi_1(\textbf{k})},\ket{\phi_2(\textbf{k})}$. Because these two vectors are kernel vectors of $A^t(\textbf{k})A(\textbf{k})$, the quantum metric and the Euler curvature become the same with the two zero modes of $A^t(\textbf{k})A(\textbf{k})$. Since $A^t(\textbf{k})A(\textbf{k})$ is smaller than $H(\textbf{k})$, in this case, due to the symmetry constraint, there exists the system with a smaller number of bands with the same band geometry.

When $M=N$, the two middle bands are a chiral symmetric pair with respect to each other. In this case, since the wavefunctions are real, the Euler curvature between the middle two bands can be written as 
\begin{align}
    \begin{split}
    F_{12}(\textbf{k})&=\bra{\partial_x u(\textbf{k})}S\ket{\partial_y u(\textbf{k})}-\bra{\partial_y u(\textbf{k})}S\ket{\partial_x u(\textbf{k})}\\
    &=\bra{\partial_x u(\textbf{k})}S\ket{\partial_y u(\textbf{k})}-\bra{\partial_x u(\textbf{k})}S^t\ket{\partial_y u(\textbf{k})},
    \end{split}
\end{align}
where $\ket{u(k)}$ is a Bloch wavefunction for one of the two middle bands. Since the real chiral symmetry operator with $S^2=1$ is symmetric, this becomes zero.

\subsection{Chiral symmetry $S^2=1$ that anti-commutes with $I_{ST}^2=1$}\label{chiralsection}
For the real gauge, the chiral symmetry that anti-commutes with the $I_{ST}$ operator should be purely imaginary. In this case, by orthogonal transformation, $S^2=1$ is orthogonally equivalent to $\tau_y$. The real $2N$-band Bloch Hamiltonian that anti-commutes with $\tau_y$ has the form 
\begin{align}
    H(\textbf{k})=
    \begin{pmatrix}
    A(\textbf{k}) & B(\textbf{k})\\
    B(\textbf{k}) & -A(\textbf{k})\\
    \end{pmatrix},
\end{align}
where $A(\textbf{k}), B(\textbf{k})$ are real and symmetric $N\times N$ matrices. The eigenvalue equation for this Bloch Hamiltonian can be written as 
\begin{align}
    \begin{pmatrix}
    A(\textbf{k}) & B(\textbf{k})\\
    B(\textbf{k}) & -A(\textbf{k})\\
    \end{pmatrix}
    \ket{\psi_n(\textbf{k})}
    &=
    E_n(\textbf{k})\ket{\psi_n(\textbf{k})},\\
    \ket{\psi_n(\textbf{k})}&=
    \begin{pmatrix}
    \ket{\psi^R_n(\textbf{k})}\\
    \ket{\psi^I_n(\textbf{k})}\\
    \end{pmatrix},
\end{align}
where $\ket{\psi^R_n(\textbf{k})}$ and $\ket{\psi^I_n(\textbf{k})}$ has the same size. The eigenvalue equation is equivalent to 
\begin{align}
\begin{split}
\label{eq:eigen}
   &\left(-A-iB(\textbf{k})\right)(\ket{\psi^R_n(\textbf{k})}-i\ket{\psi^I_n(\textbf{k})})\\
   &=-E_n(\textbf{k})(\ket{\psi^R_n(\textbf{k})}+i\ket{\psi^I_n(\textbf{k})}). 
\end{split}
\end{align}
This equation is related to the singular value decomposition of $-A(\textbf{k})-iB(\textbf{k})$. Since $-A(\textbf{k})-iB(\textbf{k})$ is a complex symmetric matrix, the singular value decomposition has a form $-A(\textbf{k})-iB(\textbf{k})=U(\textbf{k})\Lambda(\textbf{k})U^t(\textbf{k})$, where $U(\textbf{k})$ is a unitary matrix and $\Lambda(\textbf{k})$ is a diagonal matrix having semi-positive elements. The diagonal elements of $\Lambda(\textbf{k})$ are called the singular values, and the column vectors of the unitary matrix are called the singular vectors. For nth singular vector $\ket{\phi_n(\textbf{k})}$,
\begin{align}
\label{eq:singular}
(-A(\textbf{k})-iB(\textbf{k}))(\ket{\phi_n(\textbf{k})})^*=s_n(\textbf{k})\ket{\phi_n(\textbf{k})},
\end{align}
where $s_n(\textbf{k})$ is the nth singular value. The Eq.~(\ref{eq:eigen}) and Eq.~(\ref{eq:singular}) is identical if 
\begin{align}
\ket{\psi_n^R(\textbf{k})}&=\Re\left[\ket{\phi_n(\textbf{k})}\right],\\
\ket{\psi_n^I(\textbf{k})}&=\Im\left[\ket{\phi_n(\textbf{k})}\right],\\ E_n(\textbf{k})&=-s_n(\textbf{k}).
\end{align}
By this correspondence, the eigenvalues and eigenstates for half of the Bloch bands with semi-negative energy can be obtained from singular values and singular vectors. The remaining half of the Bloch bands with semi-positive energy can be obtained by
\begin{align}
\ket{\psi_{2N-n+1}^R(\textbf{k})}&=\Re\left[i\ket{\phi_n(\textbf{k})}\right],\\
\ket{\psi_{2N-n+1}^I(\textbf{k})}&=\Im\left[i\ket{\phi_n(\textbf{k})}\right],\\ E_{2N-n+1}(\textbf{k})&=s_n(\textbf{k}).
\end{align}

According to these correspondences, the wavefunction of the middle two bands can be represented by the $N$th singular vector as
\begin{align}
\label{eq:middletwo}
    \ket{\psi_N(\textbf{k})}=
    \begin{pmatrix}
    \Re[\ket{\phi_N(\textbf{k})}]\\
    \Im[\ket{\phi_N(\textbf{k})}]
    \end{pmatrix},
    \ket{\psi_{N+1}(\textbf{k})}=
    \begin{pmatrix}
    \Re[i\ket{\phi_N(\textbf{k})}]\\
    \Im[i\ket{\phi_N(\textbf{k})}]
    \end{pmatrix}.
\end{align}
From Eq~(\ref{eq:middletwo}), the Euler curvature between the middle two states 
\begin{align}
\label{eq:BerryEuler}
\begin{split}
&\bra{\partial_x\psi_N(\textbf{k})}\ket{\partial_y\psi_{N+1}(\textbf{k})}-\bra{\partial_y\psi_N(\textbf{k})}\ket{\partial_x\psi_{N+1}(\textbf{k})}\\
=&\Re[i\left(\bra{\partial_x\phi_N(\textbf{k})}\ket{\partial_y\phi_N(\textbf{k})}-\bra{\partial_y\phi_N(\textbf{k})}\ket{\partial_x\phi_N(\textbf{k})}\right)]\\
=&i\left(\bra{\partial_x\phi_N(\textbf{k})}\ket{\partial_y\phi_N(\textbf{k})}-\bra{\partial_y\phi_N(\textbf{k})}\ket{\partial_x\phi_N(\textbf{k})}\right)
\end{split}
\end{align}

is the same as the Berry curvature of $\ket{\phi_N(\textbf{k})}$. For the two bands in the middle to be isolated, the smallest singular value $s_N(\textbf{k})$ should not be degenerate in the whole Brillouin zone. This means the corresponding singular vector $\ket{\phi_N(\textbf{k})}$ is smooth over the Brillouin zone. Since there is no gauge freedom for $\ket{\phi_N(\textbf{k})}$, the Chern number should be zero. Therefore, by Eq.~(\ref{eq:BerryEuler}) the Euler number of the middle two bands should be zero. 

The fact that the isolated band for the smooth symmetric complex matrix can not have a nontrivial Chern number can also be seen in a less rigorous but more persuasive way. In the case of a Hermitian matrix, the Hermitian matrix can be smooth even when the phase of the unitary matrix in the decomposition $H=UDU^{\dagger}$ is not smooth, because the phase factor in the unitary matrix does not change the Hermitian matrix. However, because the complex symmetric matrix is decomposed as $UDU^t$, the additional phase factor in $U$ changes the complex symmetric matrix. Therefore, when the phase in the unitary matrix is not smooth, the complex symmetric matrix also can not be smooth. Thus, the singular vector for the smooth complex symmetric matrix should have a smooth phase, which does not allow a nontrivial Chern number.

\subsection{The mirror symmetry $M_z$ in spinful system}
In a spinful system, the mirror symmetry $M_z:(x,y,z)\rightarrow(x,y,-z)$ which the mirror plane is parallel to the system commutes with $I_{ST}=C_{2z}T$. Therefore, for $M_z=i\sigma_z$, the $C_{2z}T$ operator can be represented as
\begin{align}
    I_{ST}=
    \begin{pmatrix}
        0 & U_1 \\
        U_2 & 0
    \end{pmatrix}
    K,
\end{align}
where $U_1, U_2$ are unitary matrices satisfying $U_1U_2^*=U_2U_1^*=1$ because of $I_{ST}^2=1$. $I_{ST}$ can become $\sigma_x K$ by applying a unitary transformation mixing states with mirror eigenvalue $-i$.
\begin{align}
    \begin{split}
    &\begin{pmatrix}
    1 & 0\\
    0 & U_2^{-1}\\
    \end{pmatrix}
    I_{ST}
    \begin{pmatrix}
    1 & 0\\
    0 & U_2\\
    \end{pmatrix}\\
    &=
    \begin{pmatrix}
    1 & 0\\
    0 & U_2^{-1}\\
    \end{pmatrix}
    \begin{pmatrix}
    0 & U_1K\\
    U_2K & 0\\
    \end{pmatrix}
    \begin{pmatrix}
    1 & 0\\
    0 & U_2\\
    \end{pmatrix}
    =
    \begin{pmatrix}
    0 & 1\\
    1 & 0\\
    \end{pmatrix}
    K=\sigma_x K,
    \end{split}
\end{align}

Under mirror symmetry, Hamiltonian can be written as 
\begin{align}
    H(\textbf{k})=
    \begin{pmatrix}
        H_{+}(\textbf{k}) & 0\\
        0 & H_{-}(\textbf{k})\\
    \end{pmatrix},
\end{align}
where $H_{+}(\textbf{k})$ couples states with mirror eigenvalue $i$ , and $H_{-}(\textbf{k})$ couples states with mirror eigenvalue $-i$. Also with additional $I_{ST}=C_{2z}T=\sigma_x K$ symmetry, $H_{+}(\textbf{k})=H_{-}(\textbf{k})^*$ holds. The related mapping from the Chern insulator to the Euler insulator is explained in the next section.

\subsection{Mapping from $N$-band the Chern insulator to $2N$-band the Euler insulator}
Regardless of the total band number of the Chern insulator, the Chern insulator having an isolated nontrivial band can be mapped into the Euler insulator by combining the Chern insulator with its $I_{ST}$ symmetric pair. The relationship between the Hamiltonian for the Euler insulator and the Chern insulator is given by
\begin{align}
  H_{Euler}(\textbf{k})= 
\begin{pmatrix}
H_{Chern}(\textbf{k}) & 0\\
0 & H_{Chern}^*(\textbf{k})\\
\end{pmatrix},
\end{align}

where $H_{Chern}(\textbf{k})$ is the Bloch Hamiltonian for the Chern insulator. To obtain the Euler curvature for this model we should change it to the real gauge. It can be done by the unitary transformation 
\begin{align}
U=
\begin{pmatrix}
\frac{1}{\sqrt{2}}I_n & \frac{1}{\sqrt{2}}I_n\\
-\frac{i}{\sqrt{2}}I_n & \frac{i}{\sqrt{2}}I_n\\
\end{pmatrix}.
\end{align}
 This transforms $H_{Euler}$ into

\begin{align}
\label{eq:EulerChern}
H_{Euler}(\textbf{k})= 
\begin{pmatrix}
\Re[H_{Chern}(\textbf{k})] & -\Im[H_{Chern}(\textbf{k})]\\
\Im[H_{Chern}(\textbf{k})] & \Re[H_{Chern}(\textbf{k})]\\
\end{pmatrix}.
\end{align}

For real gauge, real eigenstates have the following correspondence with eigenstates of $H_{Chern}$.
\begin{align}
\ket{\phi_n^A(\textbf{k})}=
\begin{pmatrix}
\Re[\ket{u_n(\textbf{k})}]\\
\Im[\ket{u_n(\textbf{k})}]
\end{pmatrix},\\ 
\ket{\phi_n^B(\textbf{k})}=
\begin{pmatrix}
-\Im[\ket{u_n(\textbf{k})}]\\
\Re[\ket{u_n(\textbf{k})}]
\end{pmatrix},
\end{align}
where $\ket{u_n(\textbf{k})}$ is nth eigenstate of $H_{Chern}(\textbf{k})$ and $\ket{\phi^{A,B}_n(\textbf{k})}$ are corresponding eigenstates for $H_{Euler}(\textbf{k})$. The eigenvalue of real eigenstates are same with eigenvalue of $\ket{u_n(\textbf{k})}$. Therefore, if $\ket{u_n(\textbf{k})}$ is isolated band in $H_{Chern}(\textbf{k})$, $\ket{\phi_n^A(\textbf{k})}$ and $\ket{\phi_n^B(\textbf{k})}$ becomes two isolated degenerate bands in the mapped insulator.

From these correspondences, the off-diagonal term of the real Berry curvature of two isolated degenerate bands
\begin{align}
\bra{\partial_x\phi_n^A(\textbf{k})}\ket{\partial_y\phi_n^B(\textbf{k})}-\bra{\partial_y\phi_n^B(\textbf{k})}\ket{\partial_x\phi_n^A(\textbf{k})},
\end{align}
becomes the same as the Berry curvature of $\ket{u_n(\textbf{k})}$
\begin{align}
i(\bra{\partial_x u_n(\textbf{k})}\ket{\partial_y u_n(\textbf{k})}-\bra{\partial_y u_n(\textbf{k})}\ket{\partial_x u_n(\textbf{k})}).
\end{align}
 As a result, the Euler class for the two bands becomes identical to the Chern number of $\ket{u_n(\textbf{k})}$.
Also, each element of the metric 
\begin{align}
\sum_{\lambda=A,B}\bra{\partial_i \phi_n^\lambda(\textbf{k})}(1-P(\textbf{k})_{AB})\ket{\partial_j \phi_n^\lambda(\textbf{k})},
\end{align}
where $P_{AB}=\ket{\phi_n^A(\textbf{k})}\bra{\phi_n^A(\textbf{k})}+\ket{\phi_n^B(\textbf{k})}\bra{\phi_n^B(\textbf{k})}$ is a projector onto the two bands, is twice that of 
\begin{align}
\bra{\partial_i u_n(\textbf{k})}(1-P(\textbf{k}))\ket{\partial_j u_n(\textbf{k})}.    
\end{align}
Because every element of the metric is doubled, $\sqrt{\det(g_{\mu\nu}(\textbf{k}))}$ is also doubled. Therefore, the quantum volume becomes doubled. As a result, the inequality Eq. (\ref{eq:Inequality}) saturates for $\ket{\phi_n^{A,B}(\textbf{k})}$ if and only if the inequality
\begin{align}
\label{eq:inequalityChern}
    \sqrt{\det(g_{\mu\nu}(\textbf{k}))}\geq \frac{1}{2}|\Omega(\textbf{k})|
\end{align}
given in \cite{chernAndVolume} saturates for $\ket{u_n(\textbf{k})}$

\subsection{Example of four-band the Euler insulator}
In this section, we construct a four-band model that can be mapped to two superposed two-band Chern insulators in certain limits. In such limits, as the inequality in Eq.~(\ref{eq:inequalityChern}) becomes the equality for two-band Chern insulators, a similar equality should hold for Euler bands.
Explicitly, we consider the Hamiltonian $H^{a}_{4}(\textbf{k})$ with components
$H^{a}_{4}(\textbf{k})=a(\textbf{k})\sigma_x+b(\textbf{k})\tau_y\otimes\sigma_y+c(\textbf{k})\sigma_z+m_3\tau_x\otimes\sigma_x$ where $a(\textbf{k}),b(\textbf{k}),c(\textbf{k})$ are the same as those in the previous three-band model with $t=1$ and $m_1=m_2=0$. 
The band dispersion at $m_3=0$ is shown in Fig.~\ref{fig:fourBand}(a).
When $m_3=0$, the Hamiltonian commutes with $\tau_y$, and thus can be written as in Eq.~(\ref{eq:EulerChern}). 
As shown in Fig.~\ref{fig:fourBand}(c),  $vol_g/2\pi$ and $|e_2|$ of lower degenerate bands coincide only at $m_3=0$. The same also holds for upper degenerate bands.

We also constructed another four-band real Hamiltonian having chiral symmetry $S^2=1$, which satisfies $[S,I_{ST}]=0$ in certain limits. Explicitly, we consider the Hamiltonian $H^{b}_{4}(\textbf{k})$ with components
$H^{b}_{4}(\textbf{k})_{ij}=v_i(\textbf{k})v_j(\textbf{k})-w_i(\textbf{k})w_j(\textbf{k})+m_3\sigma_x\otimes\tau_x$ in which
$v^T(\textbf{k})=[1,a(\textbf{k}),b(\textbf{k}),c(\textbf{k})]$,
$w^T(\textbf{k})=[-1,a(\textbf{k}),b(\textbf{k}),c(\textbf{k})]$
where $a(\textbf{k}),b(\textbf{k}),c(\textbf{k})$ are the same as those in $H_4^a(\textbf{k})$. 
$H^{b}_{4}(\textbf{k})$ is chiral symmetric with respect to $diag(1,-1,-1,-1)$ only at $m_3=0$. 

\begin{figure}
	\includegraphics[width=0.45\textwidth]{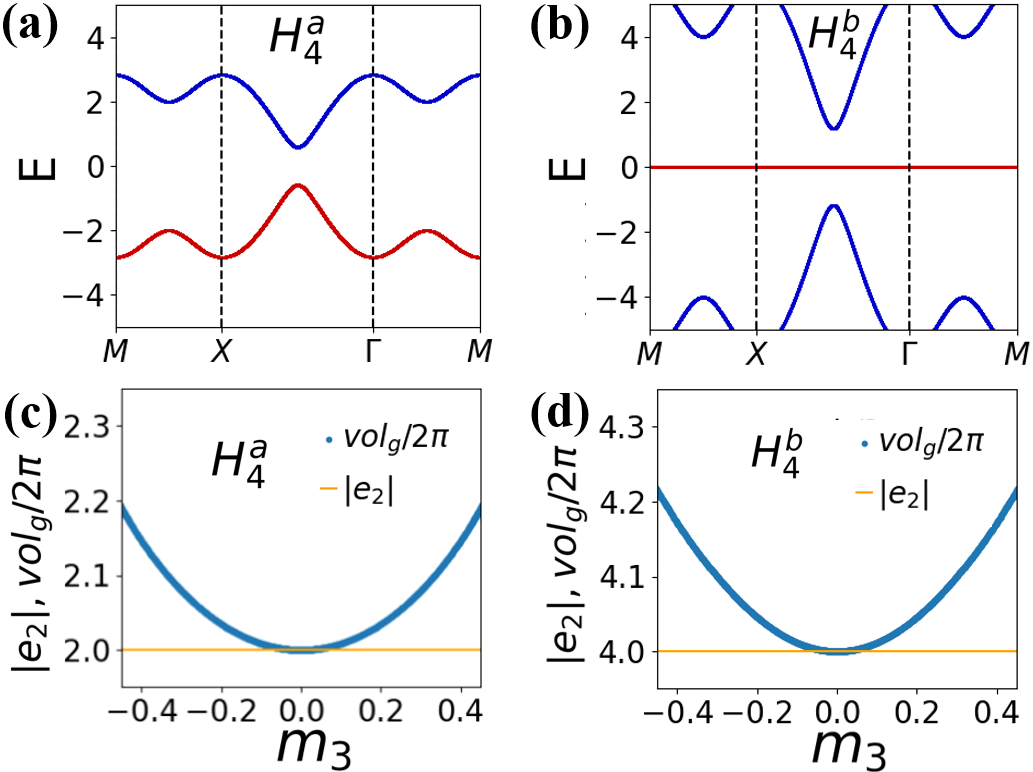}
 \vspace{-0.3cm}
	\caption{(a) The band structure of the four-band model $H^{a}_{4}(\textbf{k})$ when $m_3=0$. Both the red and blue bands are doubly degenerate when $m_3=0$ while the degeneracy is lifted when $m_3\neq 0$. (b) Similar plot for $H^{b}_{4}(\textbf{k})$ when $m_3=0$. Only the red line is doubly degenerate.
	(c) The quantum volume and Euler invariant of the red band in (a). (d) The quantum volume and Euler invariant of the red band in (b).	
}
	\label{fig:fourBand}
\end{figure}

\section{Ideal condition}
For the flat Chern bands, by numerical calculations~\cite{Ideal_one}, it was shown that the many-body ground states with fractional topology are robust when $\Omega_{xy}(\textbf{k})\neq 0$ and 
\begin{align}
\label{eq:ChernIdealCondition}
    g_{\mu\nu}(\textbf{k})=\frac{1}{2}\Omega_{xy}(\textbf{k})\omega_{\mu\nu},
\end{align}
where the determinant of $\omega_{\mu\nu}$ is one. This condition is named as the ideal condition in \cite{Ideal_one}. We suggest that the corresponding condition for the Euler insulator is given as $F_{12}(\textbf{k})\neq0$ and 
\begin{align}
\label{eq:EulerIdealCondition}
    g_{\mu\nu}(\textbf{k})=F_{12}(\textbf{k})\omega_{\mu\nu},
\end{align}
where the determinant of $\omega_{\mu\nu}$ is one. We call it the ideal condition for the Euler band. In this section, we show that this condition has various characteristics which are similar to the ideal condition of a Chern band.

\subsection{Relation with the inequality}
When the ideal condition holds, the inequality between the Berry curvature and quantum metric, Eq.(\ref{eq:inequalityChern}), Eq.(\ref{eq:Inequality}), saturates because, Eq.(\ref{eq:ChernIdealCondition}), Eq.(\ref{eq:EulerIdealCondition}), holds. Therefore, the ideal condition is a stronger condition than the inequality. In this section, we find a condition that becomes an ideal condition when combined with the saturation of inequality. For the Chern insulator, we follow the approach in \cite{Ideal_one}. 

For one isolated Chern band the inequality
\begin{align}
\label{eq:InequalityChern2}
    \sqrt{det(g_{\mu\nu}(\textbf{k}))}\geq\frac{1}{2}\left|\Omega(\textbf{k})\right|
\end{align}
saturates when the determinant of
\begin{align}
\label{eq:contractedQGT}
    Q_{\mu\nu}(\textbf{k})=g_{\mu\nu}(\textbf{k})-i\epsilon_{\mu\nu}\Omega(\textbf{k})/2,
\end{align}
which is the QGT given in Eq. (\ref{eq:QGT}), is zero. Therefore, when the inequality saturates $Q_{\mu\nu}$ has a null vector $\lambda(\textbf{k})$,
\begin{align}
\label{eq:null}
    \sum_{\nu=x,y} Q_{\mu\nu}(\textbf{k})\lambda_\nu(\textbf{k})=0.
\end{align}
In this case, $Q_{\mu\nu}(\textbf{k})$ is $2\times2$ Hermitian matrix with rank one. Since the column vectors of rank one $2\times2$ matrix should be linearly dependent
\begin{align}
    &Q_{\mu\nu}(\textbf{k})\propto \bar{\rho}_\nu(\textbf{k})\bar{\lambda}_\mu(\textbf{k})^*,
\end{align}
where $\bar{\rho}_\nu(\textbf{k})$ is complex vector and $\bar{\lambda}_\nu(\textbf{k})$ is complex vector orthogonal to $\lambda_\nu(\textbf{k})$.
\begin{align}
&\sum_{\mu=x,y}\bar{\lambda}_\mu(\textbf{k})^*\lambda_\mu(\textbf{k})=0.    
\end{align}
Bcause $Q_{\mu\nu}(\textbf{k})$ is Hermitian,
\begin{align}
    &Q_{\mu\nu}(\textbf{k})\propto \bar{\lambda}_\mu(\textbf{k})\bar{\lambda}_\nu(\textbf{k})^*.
\end{align}

By choosing the appropriate normalization for $\bar{\lambda}_{\mu}(\textbf{k})$, the quantum geometric tensor can be written using the Berry curvature and the null vector.
\begin{align}
\label{eq:QGTnull}
    &Q_{\mu\nu}(\textbf{k})=|\Omega(\textbf{k})|\bar{\lambda}_\mu(\textbf{k})\bar{\lambda}_\nu(\textbf{k})^*,
\end{align}
The normalization condition is
\begin{align}
\label{eq:normalization}
    -i\epsilon_{\mu\nu}\frac{\Omega(\textbf{k})}{|\Omega(\textbf{k})|}=\bar{\lambda}_\mu(\textbf{k})\bar{\lambda}_\nu(\textbf{k})^*-\bar{\lambda}_\mu(\textbf{k})^*\bar{\lambda}_\nu(\textbf{k}).
\end{align}
Here we note that 
\begin{align}
    &Q_{\mu\nu}(\textbf{k})=\Omega(\textbf{k})\bar{\lambda}_\mu(\textbf{k})\bar{\lambda}_\nu(\textbf{k})^*,
\end{align}
is not possible when the Berry curvature is negative, because from 
\begin{align}
    Q_{xx}&=\bra{\partial_xu_n(\textbf{k})}\left(1-P(\textbf{k})\right)\ket{\partial_xu_n(\textbf{k})}\\
    &=\left|\left(1-P(\textbf{k})\right)\ket{\partial_xu_n(\textbf{k})}\right|^2\geq 0,\\
    Q_{yy}&=\bra{\partial_yu_n(\textbf{k})}\left(1-P(\textbf{k})\right)\ket{\partial_yu_n(\textbf{k})}\\
    &=\left|\left(1-P(\textbf{k})\right)\ket{\partial_yu_n(\textbf{k})}\right|^2\geq 0,
\end{align}
the diagonal components of the quantum geometric tensor are not negative.

From Eq. (\ref{eq:contractedQGT}), Eq. (\ref{eq:QGTnull}) the metric can be represented as
\begin{align}
    g_{\mu\nu}(\textbf{k})&=\frac{1}{2}|\Omega(\textbf{k})|(\bar{\lambda}_\mu(\textbf{k})\bar{\lambda}_\nu(\textbf{k})^*+\bar{\lambda}_\mu(\textbf{k})^*\bar{\lambda}_\nu(\textbf{k}))\\
    &=\frac{1}{2}\Omega(\textbf{k})\omega_{\mu\nu}(\textbf{k}),\\
    \omega_{\mu\nu}(\textbf{k})&=\frac{\Omega(\textbf{k})}{|\Omega(\textbf{k})|}(\bar{\lambda}_\mu(\textbf{k})\bar{\lambda}_\nu(\textbf{k})^*+\bar{\lambda}_\mu(\textbf{k})^*\bar{\lambda}_\nu(\textbf{k}))
\end{align}
The determinant of $\omega_{\mu\nu}(\textbf{k})$ is one because, from Eq. (\ref{eq:normalization}),
\begin{align}
    \begin{split}
    &\det(\omega_{\mu\nu}(\textbf{k}))=\left(\frac{\Omega(\textbf{k})}{|\Omega(\textbf{k})|}\right)^2\det(\lambda_\alpha^*\lambda_\beta+\lambda_\beta^*\lambda_\alpha)\\
    &=(2|\lambda_x|^2)(2|\lambda_y|^2)-|\lambda_x^*\lambda_y+\lambda_x\lambda_y^*|^2\\
    &=2|\lambda_x|^2|\lambda_y|^2-(\lambda_x^*\lambda_y)^2-(\lambda_x\lambda_y^*)^2\\
    &=|\lambda_x^*\lambda_y-\lambda_x\lambda_y^*|^2=1.
    \end{split}
\end{align}
To be ideal the only condition needed is that $\omega_{\mu\nu}(\textbf{k})$ is constant. For $\omega_{\mu\nu}(\textbf{k})$ to be well defined, the Berry curvature should not be zero, and for $\omega_{\mu\nu}(\textbf{k})$ to be constant the null vector should be constant also the sign of the Berry curvature should be fixed. Therefore, the additional conditions needed for ideal other than the saturation of the inequality Eq.(\ref{eq:InequalityChern2}) is that the Berry curvature should be nonzero and the null vector for the QGT should be constant.

For the Euler insulator the inequality
\begin{align}
    \sqrt{det(g_{\mu\nu}(\textbf{k}))}\geq\left|F_{12}(\textbf{k})\right|
\end{align}
saturates when the determinant of the following matrix is zero.
\begin{align}
    G_{\mu\nu}(\textbf{k})=g_{\mu\nu}(\textbf{k})+i\epsilon_{\mu\nu}F_{12}(\textbf{k})
\end{align}
Therefore, $G_{\mu\nu}(\textbf{k})$ in the Euler bands corresponds to $Q_{\mu\nu}(\textbf{k})$ in the Chern band. Following the same logic in the Chern insulator, the Euler bands become ideal if and only if the inequality saturates and the null vector for the $G_{\mu\nu}(\textbf{k})$ becomes constant and the Euler curvature does not become zero for every $\textbf{k}$, as in a Chern band.

\subsection{Relation with holomorphicity}
The constant null vector in the previous section is related to the holomorphicity of the Bloch wave function. For the Chern insulator we follow the logic in \cite{Ideal_one}. The QGT of the ideal band in the Chern insulator has a constant null vector $\lambda_{\mu}$,
\begin{align}
    \sum_\nu Q_{\mu\nu}(\textbf{k})\lambda_{\nu}=0.
\end{align}
From the definition of QGT in Eq(\ref{eq:QGT}),
\begin{align}
\begin{split}
    &0=\sum_{\mu\nu} Q_{\mu\nu}(\textbf{k})\lambda_{\mu}^*\lambda_{\nu}\\
    &=\sum_{\mu\nu} \bra{\partial_\mu u_n(\textbf{k})}(1-\ket{u_n(\textbf{k})}\bra{u_n(\textbf{k})})\ket{\partial_\nu u_n(\textbf{k})}\lambda_{\mu}^*\lambda_{\nu}\\
    &=\bra{\partial_{k^*} u_n(\textbf{k})}(1-\ket{u_n(\textbf{k})}\bra{u_n(\textbf{k})})\ket{\partial_{k^*} u_n(\textbf{k})},
\end{split}\\
    &k=\lambda_x^*k_x+\lambda_y^*k_y,\\
    &\therefore (1-\ket{u_n(\textbf{k})}\bra{u_n(\textbf{k})})\ket{\partial_{k^*} u_n(\textbf{k})}=0.
\end{align}
The derivative of the Bloch wavefunction by a complex number $k=\lambda_x^*k_x+\lambda_y^*k_y$ is parallel to the Bloch wavefunction. Therefore the complex function $\lambda(\textbf{k})$
\begin{align}
\label{eq:nearHolo}
    \ket{\partial_{k^*} u_n(\textbf{k})}=\lambda(\textbf{k})\ket{u_n(\textbf{k})},
\end{align}
exists. If the derivative of the Bloch wavefunction by $k^*$ is zero, it is a holomorphic function of $k$. Since the derivative is parallel to the Bloch wavefunction, the derivative can be zero by multiplying the Bloch wavefunction by the appropriate complex function. By Eq. (\ref{eq:nearHolo}), this complex function can be found by solving,

\begin{align}
\label{eq:holo}
0=\partial_{k^*}\left(C_{\textbf{k}}\ket{u_n(\textbf{k})}\right)=\left(\partial_{k^*}C_{\textbf{k}}+C_{\textbf{k}}\lambda(\textbf{k})\right)\ket{u_n(\textbf{k})}.
\end{align}

From Eq. (\ref{eq:holo}), $C_\textbf{k}\ket{ u_n(\textbf{k})}$ is holomorphic function of $k$.
\begin{align}
\label{eq:holodead}
    C_\textbf{k}\ket{ u_n(\textbf{k})}=\ket{\tilde{u}_n(k)}.
\end{align}
The phase factor can be absorbed to $\ket{ u_n(\textbf{k})}$ because the Bloch wavefunction has a phase degree of freedom. Therefore, the real function $N_\textbf{k}$ can be used instead of the complex function.
\begin{align}
\label{eq:ChernIdealholo}
    \ket{ u_n(\textbf{k})}=\frac{1}{N_\textbf{k}}\ket{\tilde{u}_n(k)}.
\end{align}
Therefore, an ideal Chern band can be decomposed into a 
holomorphic function of $k=\lambda_x^*k_x+\lambda_y^*k_y$ and the normalization factor, where $\lambda_\mu$ is the null vector of the QGT.

To find a corresponding characteristic for Euler bands, we use the relation between the QGT of the Chern basis, $\ket{u_\pm(\textbf{k})}=\left(\ket{u_1(\textbf{k})}\pm i\ket{u_2(\textbf{k})}\right)/\sqrt{2}$, and $G_{\mu\nu}(\textbf{k})$. The relation is given as
\begin{align}
\label{eq:connectingGeometry}
\begin{split}
    G_{\mu\nu}(\textbf{k})&=2g^-_{\mu\nu}(\textbf{k})+i\epsilon_{\mu\nu}\Omega^-_{xy}(\textbf{k})=2Q^-_{\mu\nu}(\textbf{k})^*\\
    &=2g^+_{\mu\nu}(\textbf{k})-i\epsilon_{\mu\nu}\Omega^+_{xy}(\textbf{k})=2Q^+_{\mu\nu}(\textbf{k}),
\end{split}
\end{align}
where $g^{\pm}_{\mu\nu}(\textbf{k})$, $\Omega^{\pm}(\textbf{k})$, and $Q^{\pm}(\textbf{k})$ are quantum metric, Berry curvature, and QGT for the Chern basis, $\ket{u_{\pm}(\textbf{k})}$. This relation holds because $\ket{u_1(\textbf{k})}$, $\ket{u_2(\textbf{k})}$ are real. A more specific proof is as follows.
For real bands, $\bra{\partial_{\mu} u_i}\ket{u_i}=0$ and $\bra{\partial_{\mu} u_i}\ket{\partial_\nu u_j}=\bra{\partial_\nu u_j}\ket{\partial_{\mu} u_i}$. Accordingly, the Euler curvature of $\ket{u_1(\textbf{k})}$, $\ket{u_2(\textbf{k})}$ has relation with Berry curvature of $\ket{u_\pm(\textbf{k})}$ as
\begin{align}
\label{eq:connectingCurvature}
\begin{split}
    &\Omega^-_{xy}(\textbf{k})=i(\bra{\partial_xu_-(\textbf{k})}\ket{\partial_yu_-(\textbf{k})}-\bra{\partial_yu_-(\textbf{k})}\ket{\partial_xu_-(\textbf{k})})\\
    &=\frac{i}{2}(\bra{\partial_xu_1(\textbf{k})}\ket{\partial_yu_1(\textbf{k})}-\bra{\partial_yu_1(\textbf{k})}\ket{\partial_xu_1(\textbf{k})})\\
    &+\frac{1}{2}(\bra{\partial_xu_1(\textbf{k})}\ket{\partial_yu_2(\textbf{k})}-\bra{\partial_yu_1(\textbf{k})}\ket{\partial_xu_2(\textbf{k})})\\
    &-\frac{1}{2}(\bra{\partial_xu_2(\textbf{k})}\ket{\partial_yu_1(\textbf{k})}-\bra{\partial_yu_2(\textbf{k})}\ket{\partial_xu_1(\textbf{k})})\\
    &+\frac{i}{2}(\bra{\partial_xu_2(\textbf{k})}\ket{\partial_yu_2(\textbf{k})}-\bra{\partial_yu_2(\textbf{k})}\ket{\partial_xu_2(\textbf{k})})\\
    &=\bra{\partial_xu_1(\textbf{k})}\ket{\partial_yu_2(\textbf{k})}-\bra{\partial_yu_1(\textbf{k})}\ket{\partial_xu_2(\textbf{k})}=F_{12}(\textbf{k})\\
    &=-i(\bra{\partial_xu_+(\textbf{k})}\ket{\partial_yu_+(\textbf{k})}-\bra{\partial_yu_+(\textbf{k})}\ket{\partial_xu_+(\textbf{k})})=-\Omega^+_{xy}(\textbf{k}).
\end{split}
\end{align}
where $\Omega^\pm_{xy}(\textbf{k})$ are Berry curvature for $\ket{u_\pm(\textbf{k})}$. The quantum metric also has relation as
\begin{align}
\label{eq:connectingMetric}
\begin{split}
    g^+_{\mu\nu}(\textbf{k})&=\Re[\bra{\partial_\mu u_+(\textbf{k})}(1-\ket{u_+(\textbf{k})}\bra{u_+(\textbf{k})})\ket{\partial_\nu u_+(\textbf{k})}]\\
    =\frac{1}{2}\Re&[\bra{\partial_\mu u_1(\textbf{k})}\ket{\partial_\nu u_1(\textbf{k})}+\bra{\partial_\mu u_2(\textbf{k})}\ket{\partial_\nu u_2(\textbf{k})}]\\
    -\frac{1}{4}Re&\Bigl[\Bigl(\bra{\partial_\mu u_1(\textbf{k})}\ket{ u_1(\textbf{k})}+\bra{\partial_\mu u_2(\textbf{k})}\ket{ u_2(\textbf{k})}\\
    &+i\bra{\partial_\mu u_1(\textbf{k})}\ket{ u_2(\textbf{k})}-i\bra{\partial_\mu u_2(\textbf{k})}\ket{ u_1(\textbf{k})}\Bigr)\\
    &\Bigl(\bra{u_1(\textbf{k})}\ket{\partial_\nu u_1(\textbf{k})}+\bra{u_2(\textbf{k})}\ket{ \partial_\nu u_2(\textbf{k})}\\
    &+i\bra{u_1(\textbf{k})}\ket{\partial_\nu u_2(\textbf{k})}-i\bra{u_2(\textbf{k})}\ket{\partial_\nu u_1(\textbf{k})}\Bigr)\Bigr]\\
    =\frac{1}{2}\Re&[\bra{\partial_\mu u_1(\textbf{k})}\ket{\partial_\nu u_1(\textbf{k})}+\bra{\partial_\mu u_2(\textbf{k})}\ket{\partial_\nu u_2(\textbf{k})}]\\
    -\frac{1}{2}\Re&[\bra{\partial_\mu u_1(\textbf{k})}\ket{ u_2(\textbf{k})}\bra{u_2(\textbf{k})}\ket{\partial_\nu u_1(\textbf{k})}\\
    &+\bra{\partial_\mu u_2(\textbf{k})}\ket{ u_1(\textbf{k})}\bra{u_1(\textbf{k})}\ket{\partial_\nu u_2(\textbf{k})}]\\
    =\frac{1}{2}g_{\mu\nu}&(\textbf{k})=g^-_{\mu\nu}(\textbf{k}),
\end{split}
\end{align}
where $g^\pm_{xy}(\textbf{k})$ are quantum metric for $\ket{u_\pm(\textbf{k})}$. From Eq. (\ref{eq:connectingCurvature}) and Eq. (\ref{eq:connectingMetric}), Eq. (\ref{eq:connectingGeometry}) holds.

When the Euler bands are ideal, $G_{\mu\nu}(\textbf{k})$ has a constant null vector $\lambda_{\mu}$ and by Eq. (\ref{eq:connectingGeometry}), $\lambda_{\mu}$ is the null vector for QGT of $\ket{u_+(\textbf{k})}$ and $\lambda_\mu^*$ is the null vector for QGT of $\ket{u_-(\textbf{k})}$. By following the procedure in the Chern band case, we can prove that each Chern basis can be decomposed into the complex normalization factor and the holomorphic function.
\begin{align}
\label{eq:BeforeEulerIdealholo}
    &\ket{u_+(\textbf{k})}=\frac{1}{C_{\textbf{k}}}\ket{\tilde{u}(k)},\\
    &\ket{u_-(\textbf{k})}=\left(\ket{u_+(\textbf{k})}\right)^*=\frac{1}{C_{\textbf{k}}^*}\left(\ket{\tilde{u}(k)}\right)^*,\\
    &k=\lambda_x^*k_x+\lambda_y^*k_y
\end{align}
In a strict sense, the Chern basis should have a phase degree of freedom for the Chern basis to be decomposed into the holomorphic part and real normalization factor as
\begin{align}
\label{eq:EulerIdealholo}
    &\ket{u_+(\textbf{k})}=\frac{1}{\sqrt{2}}\left(\ket{u_1(\textbf{k})}+i\ket{u_2(\textbf{k})}\right)e^{i\phi_\textbf{k}}=\frac{1}{N_{\textbf{k}}}\ket{\tilde{u}(k)},\\
    &\ket{u_-(\textbf{k})}=\left(\ket{u_+(\textbf{k})}\right)^*=\frac{1}{N_{\textbf{k}}}\left(\ket{\tilde{u}(k)}\right)^*,\\
    &k=\lambda_x^*k_x+\lambda_y^*k_y,
\end{align}
where the phase factor of $C_{\textbf{k}}$ is absorbed by $e^{i\phi_\textbf{k}}$. However, this phase factor is not very important for the logic from now on, since QGT is gauge invariant. Therefore, we will ignore the phase degree of freedom to the Chern basis for simplicity.

\subsection{No-go theorem for the Ideal state}
When the Chern band is ideal, from Eq. (\ref{eq:ChernIdealholo}), the Bloch wavefunction can be decomposed into a normalization factor and holomorphic part as,
\begin{align}
    \ket{u_n(\textbf{k})}=\frac{1}{N_{\textbf{k}}}
    \begin{pmatrix}
        \tilde{u}_n(k)_1\\
        \tilde{u}_n(k)_2\\
        .\\
        .\\
        .\\
        \tilde{u}_n(k)_n\\
    \end{pmatrix},
\end{align}
where $\tilde{u}_n(k)_\alpha$ are holomorphic function of 
\begin{align}
k=\lambda_x\textbf{k}_x+\lambda_y\textbf{k}_y.
\end{align}
$\lambda_\mu^*$ is a null vector of QGT. Since $\tilde{u}_n(k)_\alpha$ are holomorphic, the boundary condition of the Bloch wavefunction $\ket{u_n(\textbf{k})}$ is important for determining the $\tilde{u}_n(k)_\alpha$. This boundary condition is given as,
\begin{align}
&u_n(\textbf{k}+\textbf{G})_{\alpha}=e^{i\phi_{\textbf{k},\textbf{G}}-i\textbf{G}\cdot\textbf{x}_\alpha}u_n(\textbf{k})_{\alpha},
\end{align}
where $\textbf{G}$ is a reciprocal lattice vector and $\textbf{x}_\alpha$ is a position of the orbital $\alpha$. According to \cite{Ideal_one}, for the fixed $\alpha$, $\tilde{u}_n(k)_\alpha$ is determined, except for multiplying by a constant factor, by the Chern number and $e^{i\textbf{b}_1\cdot\textbf{x}_\alpha},e^{i\textbf{b}_2\cdot\textbf{x}_\alpha}$, where $\textbf{b}_1,\textbf{b}_2$ are two reciprocal lattice vectors which are not parallel to each other.
If $e^{i\textbf{b}_1\cdot\textbf{x}_\alpha}=e^{i\textbf{b}_2\cdot\textbf{x}_\alpha}=1$ regardless of $\alpha$, the holomorphic part does not depend on $\alpha$. Therefore, the Bloch state can be written as,
\begin{align}
    \ket{u_n(\textbf{k})}=\frac{1}{N_{\textbf{k}}}
    \begin{pmatrix}
        c_1\\
        c_2\\
        .\\
        .\\
        .\\
        c_n\\
    \end{pmatrix}\tilde{u}_n(k).
\end{align}
Removing the holomorphic part doesn't change the state, so the Bloch state becomes constant. As a result, the topology should be trivial. If $\textbf{b}_1\cdot\textbf{x}_\alpha/2\pi,\textbf{b}_2\cdot\textbf{x}_\alpha/2\pi$ are rational similar logic can be used. When they are rational, there exists a natural number $N$ s.t. $e^{iN\textbf{b}_1\cdot\textbf{x}_\alpha}=e^{iN\textbf{b}_2\cdot\textbf{x}_\alpha}=1$ regardless of $\alpha$. Since $N\textbf{b}_1, N\textbf{b}_2$ are also reciprocal vectors, again the holomorphic part becomes the same.

For the Euler insulator, if $\textbf{b}_1\cdot\textbf{x}_\alpha/2\pi,\textbf{b}_2\cdot\textbf{x}_\alpha/2\pi$ are rational, by Eq. (\ref{eq:EulerIdealholo}), the Chern number of the Chern basis becomes trivial. Additionally, from Eq.~(\ref{eq:connectingCurvature}), the Euler class topology becomes trivial. Therefore, if $\textbf{x}_\alpha\cdot\textbf{G}/2\pi$ is rational for all the orbital positions $\textbf{x}_\alpha$ and reciprocal lattice vector $\textbf{G}$, then the lattice model, whether a Chern insulator or an Euler insulator, cannot have an ideal band with a nontrivial topology.

\subsection{Relation with the Lowest Landau level}
The energy spectrum of the interacting Hamiltonian of the various flat ideal bands can be obtained by applying various potentials to the Lowest Landau Level(LLL)~\cite{Ideal, Ideal_one}. To show this we should write the interacting Hamiltonian explicitly. If a band is flat, the Hamiltonian for electrons only has the interacting part. Additionally, if the band gap is larger than the interaction, mixing between occupied and unoccupied bands can be neglected. In this case, the low-energy effective Hamiltonian is the same with two-particle interaction projected to a band.
\begin{align}
&\int d\textbf{k}_2'd\textbf{k}_1'd\textbf{k}_2d\textbf{k}_1 H_{\textbf{k}_2',\textbf{k}_1',\textbf{k}_2,\textbf{k}_1} c_{n,\textbf{k}_2'}^{\dagger}c_{n,\textbf{k}_1'}^{\dagger}c_{n,\textbf{k}_2}c_{n,\textbf{k}_1},\\
\label{eq:interactingH}
\begin{split}
&H_{\textbf{k}_2',\textbf{k}_1',\textbf{k}_2,\textbf{k}_1}=\sum_{\textbf{R}_1,\alpha_1,\textbf{R}_2,\alpha_2}\int d\textbf{k}d\textbf{k}'\\
&e^{i\left(\textbf{k}_1-\textbf{k}_1'\right)\cdot\left(\textbf{R}_1+\textbf{x}_{\alpha_1}\right)+i\left(\textbf{k}_2-\textbf{k}_2'\right)\cdot\left(\textbf{R}_2+\textbf{x}_{\alpha_2}\right)}v(\textbf{R}_2+\textbf{x}_{\alpha_2},\textbf{R}_1+\textbf{x}_{\alpha_1})\\
&u_n(\textbf{k}_2')_{\alpha_2}^* u_n(\textbf{k}_1')_{\alpha_1}^*u_n(\textbf{k}_2)_{\alpha_2} u_n(\textbf{k}_1)_{\alpha_1},\\
\end{split}
\end{align}
where $\textbf{R}$ is a Bravais lattice vector, $\textbf{x}_\alpha$ is a position of the orbital $\alpha$ in the unit cell, and $c_{n,\textbf{k}}$ is an annihilation operator for $\ket{u_n(\textbf{k})}$. There are two ways to manipulate this Hamiltonian that gives the same energy spectrum. The first way is to change the Bloch wavefunction.
\begin{align}
     u_n(\textbf{k})_\alpha \rightarrow \frac{d_\alpha}{N_{\textbf{k}}}u_n(\textbf{k})_\alpha,
\end{align}
where $d_\alpha$ is a complex number depending on $\alpha$ and $N_{\textbf{k}}$ is a normalization factor. Another way is to change the interaction as
\begin{align}
\label{eq:potentialRule}
v(\textbf{R}_2+\textbf{x}_{\alpha_2},\textbf{R}_1+\textbf{x}_{\alpha_1})\rightarrow\frac{|d_{\alpha_1}d_{\alpha_2}|^2v(\textbf{R}_2+\textbf{x}_{\alpha_2},\textbf{R}_1+\textbf{x}_{\alpha_1})}{N_{\textbf{k}_2'}N_{\textbf{k}_1'}N_{\textbf{k}_2}N_{\textbf{k}_1}}.
\end{align}
Two manipulations give the same $H_{\textbf{k}_2',\textbf{k}_1',\textbf{k}_2,\textbf{k}_1}$ in Eq. (\ref{eq:interactingH}). 
This relation is important for an ideal Chern band. Because, if $\ket{u_n(\textbf{k})}$, $\ket{v_n(\textbf{k})}$ are both Bloch wavefunctions of the ideal Chern bands with the same orbital position, because they share the same holomorphic part, there exists $d_\alpha$, $N_\textbf{k}$ such that
\begin{align}
    v_n(\textbf{k})_\alpha = \frac{d_\alpha}{N_{\textbf{k}}}u_n(\textbf{k})_\alpha.
\end{align}
In \cite{Ideal_one}, stability of the ground state with fractional topology was explored for various ideal bands with $C=1$, by manipulating the potential by the rule in Eq. (\ref{eq:potentialRule}), while fixing the wavefunction with the wavefunction of the LLL. If the Chern number is not one, a linear superposition of the translated LLLs should be used to match the boundary condition of the Bloch wavefunction~\cite{Ideal}.

For the ideal Euler bands the rule becomes more complicated because there are two bands. The interacting Hamiltonian becomes,
\begin{align}
&\int d\textbf{k}_2'd\textbf{k}_1'd\textbf{k}_2d\textbf{k}_1 H_{\textbf{k}_2',\textbf{k}_1',\textbf{k}_2,\textbf{k}_1}^{n_2',n_1',n_2,n_1} c_{n_2',\textbf{k}_2'}^{\dagger}c_{n_1',\textbf{k}_1'}^{\dagger}c_{n_2,\textbf{k}_2}c_{n_1,\textbf{k}_1},\\
\label{eq:interactingH2}
\begin{split}
&H_{\textbf{k}_2',\textbf{k}_1',\textbf{k}_2,\textbf{k}_1}^{n_2',n_1',n_2,n_1}=\sum_{\textbf{R}_1,\alpha_1,\textbf{R}_2,\alpha_2}\int d\textbf{k}d\textbf{k}'\\
&e^{i\left(\textbf{k}_1-\textbf{k}_1'\right)\cdot\left(\textbf{R}_1+\textbf{x}_{\alpha_1}\right)+i\left(\textbf{k}_2-\textbf{k}_2'\right)\cdot\left(\textbf{R}_2+\textbf{x}_{\alpha_2}\right)}v(\textbf{R}_2+\textbf{x}_{\alpha_2},\textbf{R}_1+\textbf{x}_{\alpha_1})\\
&u_{n_2'}(\textbf{k}_2')_{\alpha_2}^* u_{n_1'}(\textbf{k}_1')_{\alpha_1}^*u_{n_2}(\textbf{k}_2)_{\alpha_2} u_{n_1}(\textbf{k}_1)_{\alpha_1},\\
\end{split}
\end{align}
where $c_{\pm,\textbf{k}}$ is an annihilation operator for the Chern basis $\ket{u_\pm(\textbf{k})}$. For the Chern basis of the ideal Euler bands in two different models with the same orbital positions, $u_\pm(\textbf{k})_\alpha$, $v_\pm(\textbf{k})_\alpha$, there exists $d_\alpha^\pm$, $N_\textbf{k}$ such that
\begin{align}
    v_\pm(\textbf{k})_\alpha = \frac{d_\alpha^\pm}{N_{\textbf{k}}}u_\pm(\textbf{k})_\alpha.
\end{align}
The rule for manipulating potential becomes, 
\begin{align}
\label{eq:potentialRule2}
\begin{split}
&v(\textbf{R}_2+\textbf{x}_{\alpha_2},\textbf{R}_1+\textbf{x}_{\alpha_1})\\
&\rightarrow\frac{\left(d_{\alpha_2}^{n_2'}\right)^*\left(d_{\alpha_1}^{n_1'}\right)^*d_{\alpha_2}^{n_2}d_{\alpha_1}^{n_1}v(\textbf{R}_2+\textbf{x}_{\alpha_2},\textbf{R}_1+\textbf{x}_{\alpha_1})}{N_{\textbf{k}_2'}N_{\textbf{k}_1'}N_{\textbf{k}_2}N_{\textbf{k}_1}}.
\end{split}
\end{align}
Therefore as in a flat ideal Chern band case, the energy spectrum of the interacting Hamiltonian of the various flat ideal Euler bands can be obtained by manipulating the potential instead of the wavefunction.

\section{Hamiltonian with ideal state}
In the previous section, we proved a no-go theorem for the ideal state in the lattice model. In this section, instead of the lattice model, we construct the examples of continuum Hamiltonians defined in the infinite 2D plane of $\textbf{k}$ having the ideal Euler bands.
\subsection{Equations for Ideal Euler bands}
\label{sec:simpleNumber}
Before making the Hamiltonian that has ideal Euler bands, we first summarize conditions for the ideal Euler band and equations that are useful for calculating Euler curvature, quantum metric, and the Euler number of the ideal Euler bands. Let, $\ket{u_1(\textbf{k})}$, $\ket{u_2(\textbf{k})}$ be the ideal Euler bands. From Eq. (\ref{eq:EulerIdealholo}), $\ket{u_+(\textbf{k})}=\frac{1}{\sqrt{2}}\left(\ket{u_1(\textbf{k})}+i\ket{u_2(\textbf{k})}\right)e^{i\phi_{\textbf{k}}}$ can be decomposed into the holomorphic part and the normalization part.
\begin{align}
    \label{eq:ChernBasis}
    \ket{u_{+}(\textbf{k})}&=\frac{1}{\sqrt{2}}\left(\ket{u_1(\textbf{k})}+i\ket{u_2(\textbf{k})}\right)e^{i\phi_{\textbf{k}}}\\
    &=\frac{1}{N_{\textbf{k}}}\ket{\tilde{u}(k)}=\frac{1}{N_{\textbf{k}}}
    \begin{pmatrix}
        \tilde{u}(k)_1\\
        \tilde{u}(k)_2\\
        .\\
        .\\
        .\\
        \tilde{u}(k)_n\\
    \end{pmatrix},
\end{align}
where $\tilde{u}(k)_1, ... , \tilde{u}(k)_n$ are the polynomials of a complex number $k=\lambda_x\textbf{k}_x+\lambda_y\textbf{k}_y$ and $\lambda_\mu^*$ is the constant null vector for QGT of $\ket{u_+(\textbf{k})}$. 

Even though the Chern basis of the ideal Euler band can be decomposed into a normalization factor and the holomorphic function, not any normalized holomorphic wavefunction can be a Chern basis of the ideal Euler band. To construct a Chern basis of the ideal Euler band from $\ket{\tilde{u}(k)}$ the following conditions are needed.
\begin{align}
\label{eq:condition2}
    &\forall k, \bra{\tilde{u}(k)}\ket{\tilde{u}(k)}\neq 0,\\ 
\label{eq:condition3}
    &(\bra{\tilde{u}(k)})^*\ket{\tilde{u}(k)}= 0,\\
\label{eq:curvatureCondition}
    &\forall k,F_{12}(\textbf{k})\neq0,\\
    &g_{\mu\nu}(\textbf{k})=F_{12}(\textbf{k})\omega_{\mu\nu}.
\end{align}
where $\omega_{\mu\nu}$ is a matrix with a unit determinant. First condition is needed for the normalization of $\ket{\tilde{u}(k)}$ and the second condition is needed for the normalization of $\ket{u_1(\textbf{k})}$ and $\ket{u_2(\textbf{k})}$. The third and fourth conditions are conditions for the ideal Euler band. Only the first three conditions are required for the holomorphic wavefunction to construct the ideal Euler band, because the fourth condition is automatically satisfied. The proof is as follows. By Eq.~(\ref{eq:connectingCurvature}), Eq.~(\ref{eq:connectingMetric}), and $k=\lambda_x\textbf{k}_x+\lambda_y\textbf{k}_y$,
\begin{align}
\label{eq:curvaturek}
&F_{12}(\textbf{k})=-\Omega^+_{xy}(\textbf{k})=i(\lambda_y^*\lambda_x-\lambda_x^*\lambda_y)\bra{v(\textbf{k})}\ket{v(\textbf{k})},\\
\label{eq:metrick}
&g_{\alpha\beta}(\textbf{k})=2g^+_{\alpha\beta}(\textbf{k})=
(\lambda_\alpha^*\lambda_\beta+\lambda_\beta^*\lambda_\alpha)\bra{v(\textbf{k})}\ket{v(\textbf{k})},\\
&\ket{v(\textbf{k})}=\frac{1}{N_{\textbf{k}}}(1-\ket{u_+(\textbf{k})}\bra{u_+(\textbf{k})})\ket{\partial_k\tilde{u}(k)}.
\end{align}
From Eq.~(\ref{eq:curvaturek}), Eq.~(\ref{eq:metrick}),
\begin{align}
    g_{\alpha\beta}(\textbf{k})/F_{12}(\textbf{k})=\frac{\lambda_\alpha^*\lambda_\beta+\lambda_\beta^*\lambda_\alpha}{i(\lambda_y^*\lambda_x-\lambda_x^*\lambda_y)}
\end{align}
is constant and
\begin{align}
    \begin{split}
    &\det(\lambda_\alpha^*\lambda_\beta+\lambda_\beta^*\lambda_\alpha)\\
    &=(2|\lambda_x|^2)(2|\lambda_y|^2)-|\lambda_x^*\lambda_y+\lambda_x\lambda_y^*|^2\\
    &=2|\lambda_x|^2|\lambda_y|^2-(\lambda_x^*\lambda_y)^2-(\lambda_x\lambda_y^*)^2\\
    &=(i(\lambda_x^*\lambda_y-\lambda_x\lambda_y^*))^2
    \end{split}\\
    &\therefore \det(\frac{\lambda_\alpha^*\lambda_\beta+\lambda_\beta^*\lambda_\alpha}{i(\lambda_y^*\lambda_x-\lambda_x^*\lambda_y)})=1.
\end{align}

The Euler number can be calculated by integrating $F_{12}(k)$ over the whole plane, but the calculation becomes too complex to apply to the examples that will be introduced. Instead, the Euler number can be calculated from the line integral of the connection. 
\begin{align}
\label{eq:connectionEuler}
    &e_2=\frac{1}{2\pi}\oint \textbf{A}_{12}(\textbf{k})\cdot d\textbf{k},\\
    &\textbf{A}_{12}(\textbf{k})=\bra{u_1(\textbf{k})}\nabla\ket{u_2(\textbf{k})},\\
    &\nabla\times\textbf{A}_{12}(\textbf{k})=F_{12}(\textbf{k}),
\end{align}
where the line integral is conducted over the circle with the center at $(k_x,k_y)=(0,0)$ when the radius $R\rightarrow\infty$. Since $\ket{u_1(\textbf{k})}, \ket{u_2(\textbf{k})}$ are real bands, the connection can also be written as
\begin{align}
\label{eq:connectionChern}
\begin{split}
    &\frac{1}{i}\bra{u_+(\textbf{k})}\nabla\ket{u_+(\textbf{k})}\\
    &=\frac{1}{2i}\Bigl(\bra{u_1(\textbf{k})}\nabla\ket{u_1(\textbf{k})}+\bra{u_2(\textbf{k})}\nabla\ket{u_2(\textbf{k})}\\
    &-i\bra{u_2(\textbf{k})}\nabla\ket{u_1(\textbf{k})}+i\bra{u_1(\textbf{k})}\nabla\ket{u_2(\textbf{k})}\Bigr)\\
    &=\textbf{A}_{12}(\textbf{k}).
\end{split}
\end{align}

From, Eq.~(\ref{eq:connectionEuler}) and Eq.~(\ref{eq:connectionChern}), $k=\lambda_x\textbf{k}_x+\lambda_y\textbf{k}_y$, the Euler number can be obtained from the integral
\begin{align}
\label{eq:e2k}
\begin{split}
    e_2&=\frac{1}{2\pi i}\oint -\nabla (\ln N_\textbf{k})+\frac{1}{N_\textbf{k}^2}\bra{\tilde{u}(k)}\nabla\ket{\tilde{u}(k)}\cdot d\textbf{k}\\
    &=\frac{1}{2\pi i} \oint \frac{1}{N_\textbf{k}^2} \bra{\tilde{u}(k)}\nabla\ket{\tilde{u}(k)}\cdot d\textbf{k}\\
    &=\frac{1}{2\pi i} \oint \frac{\bra{\tilde{u}(k)}\ket{\partial_k\tilde{u}(k)}}{\bra{\tilde{u}(k)}\ket{\tilde{u}(k)}} dk,
\end{split}
\end{align}
where the contour integral is conducted along a circle whose radius goes to infinity. The direction of the contour integral differs by the sign of
\begin{align}
    \Im[\lambda_x\lambda_y^*-\lambda_x^*\lambda_y].
\end{align}
If the sign is positive the direction of the contour integral is counter-clockwise and if the sign is negative the direction is clockwise. 

For the large $k$ limit,
\begin{align}
&\bra{\tilde{u}(k)}\ket{\tilde{u}(k)}\rightarrow d(k^*)^Dk^D,\\
&\bra{\tilde{u}(k)}\ket{\partial_k\tilde{u}(k)}\rightarrow dD(k^*)^Dk^{D-1},\\
\label{eq:largeK}
&\therefore \frac{\bra{\tilde{u}(k)}\ket{\partial_k\tilde{u}(k)}}{\bra{\tilde{u}(k)}\ket{\tilde{u}(k)}} \rightarrow \frac{D}{k},
\end{align}
where $D$ is the highest order of $k$ among $\tilde{u}(k)_1, ... , \tilde{u}(k)_n$ and $d$ is a real number.

From Eq. (\ref{eq:e2k}), Eq. (\ref{eq:largeK}),
\begin{align}
\label{eq:Relation}
    e_2=D\frac{\Im[\lambda_x\lambda_y^*-\lambda_x^*\lambda_y]}{\left|\Im[\lambda_x\lambda_y^*-\lambda_x^*\lambda_y]\right|}.
\end{align}

To make the calculation simple, when constructing the Hamiltonian we fixed $\lambda_\alpha$ as
\begin{align}
    \lambda_x=\pm\frac{1}{\sqrt{2}}, \lambda_y=\frac{i}{\sqrt{2}}.
\end{align}
Then the equation for the quantum metric, Euler curvature and Euler number is simplified as,
\begin{align}
\label{eq:HolomorphicEuler}
&F_{12}(\textbf{k})=\pm\bra{v(\textbf{k})}\ket{v(\textbf{k})},\\
&g_{\alpha\beta}(\textbf{k})=\delta_{\alpha\beta}\bra{v(\textbf{k})}\ket{v(\textbf{k})},\\
\label{eq:vforvendetta}
&\ket{v(\textbf{k})}=\frac{1}{N_{\textbf{k}}}(1-\ket{u_+(\textbf{k})}\bra{u_+(\textbf{k})})\ket{\partial_k\tilde{u}(k)},\\
&e_2=\pm D,
\end{align}
where $D$ is the highest order for $k$ among $\tilde{u}(k)_1, ... , \tilde{u}(k)_n$. Note that the sign of Euler curvature and the Euler number can flip by changing the sign of $\lambda_x$. The relation between the complex number $k$ and vector $\textbf{k}$ becomes,
\begin{align}
k=\frac{1}{\sqrt{2}}\left(\pm\textbf{k}_x+i\textbf{k}_y\right).
\end{align}

\subsection{Three band model}
To have a nontrivial Euler class topology, the model should have at least three bands.
\begin{align}
    &\frac{1}{\sqrt{2}}\left(\ket{u_1(\textbf{k})}+i\ket{u_2(\textbf{k})}\right)e^{i\phi_{\textbf{k}}}=\frac{1}{N_{\textbf{k}}}\ket{\tilde{u}(k)}=\frac{1}{N_{\textbf{k}}}
    \begin{pmatrix}
        \tilde{u}(k)_1\\
        \tilde{u}(k)_2\\
        \tilde{u}(k)_3\\
    \end{pmatrix},
\end{align}
where $\ket{u_1(\textbf{k})}$, $\ket{u_2(\textbf{k})}$ are the ideal Euler bands. From Eq.~(\ref{eq:condition2}), polynomials $\tilde{u}(k)_1,\tilde{u}(k)_2,\tilde{u}(k)_3$ should not share the root. From Eq.~(\ref{eq:condition3}), 
\begin{align}
    -\tilde{u}(k)_3^2=(\tilde{u}(k)_1+i\tilde{u}(k)_2)(\tilde{u}(k)_1-i\tilde{u}(k)_2).
\end{align}
Since $(\tilde{u}(k)_1-i\tilde{u}(k)_2)(\tilde{u}(k)_1+i\tilde{u}(k)_2)$ should be a perfect square of $iu_3(k)$ and polynomials $\tilde{u}(k)_1,\tilde{u}(k)_2,\tilde{u}(k)_3$ do not share roots, each $(\tilde{u}(k)_1-i\tilde{u}(k)_2)$, $(\tilde{u}(k)_1+i\tilde{u}(k)_2)$ is the perfect square of some polynomial of $k$. As a result, there exists polynomials $f_1(k),f_2(k)$ s.t.
\begin{align}
\label{eq:param1}
    \tilde{u}(k)_1&=f_1(k)^2-f_2(k)^2,\\
\label{eq:param2}
    \tilde{u}(k)_2&=i(f_1(k)^2+f_2(k)^2),\\
\label{eq:param3}
    \tilde{u}(k)_3&=2f_1(k)f_2(k),
\end{align}
where $f_1(k),f_2(k)$ do not share roots. The Hamiltonian having $\ket{u_1(\textbf{k})}, \ket{u_2(\textbf{k})}$ as lower two bands can be constructed by,

\begin{align}
\label{eq:generalthree}
    &H(\textbf{k})=\ket{z(k)}\bra{z(k)},\\
\label{eq:generalthreep}
    &\ket{z(k)}=
    \begin{pmatrix}
        f_1(k)^*f_2(k)+f_1(k)f_2(k)^*\\
        \frac{1}{i}(f_1(k)^*f_2(k)-f_1(k)f_2(k)^*)\\
        |f_2(k)|^2-|f_1(k)|^2
    \end{pmatrix}.
\end{align}
The Hamiltonian is real because all the elements of $\ket{z(k)}$ are real. Since $\ket{z(k)}$ is orthogonal to both $\ket{u_1(\textbf{k})},\ket{u_2(\textbf{k})}$, the energy eigenvalues for the lower two band is zero and the eigenstates for the lower two bands are $\ket{u_1(\textbf{k})},\ket{u_2(\textbf{k})}$. The energy of the third band is given as 
\begin{align}
\label{eq:threeEnergy}
\bra{z(k)}\ket{z(k)}=(|f_1(k)|^2+|f_2(k)|^2)^2.
\end{align}
Since $f_1(k), f_2(k)$ do not share roots, there is no complex number $k$ s.t. $f_1(k)=f_2(k)$. Therefore the energy of the third band is always larger than zero. Because the energy of the lower two bands is zero, the 3rd band is separated from $\ket{u_1(\textbf{k})}$ and $\ket{u_2(\textbf{k})}$.

From condition $\forall k,  F_{12}(\textbf{k})\neq 0$ given in Eq.~(\ref{eq:curvatureCondition}), the ideal Euler bands in the three-band system can only have the Euler number $\pm2$. The proof is as follows. From Eq.~(\ref{eq:HolomorphicEuler}), Eq.~(\ref{eq:vforvendetta}), Eq. (\ref{eq:param1}), Eq. (\ref{eq:param2}), Eq. (\ref{eq:param3}),

\begin{align}
&F_{12}(\textbf{k})=\pm\bra{v(\textbf{k})}\ket{v(\textbf{k})},\\
\begin{split}
\label{eq:3bandIdeal}
    &\ket{v(\textbf{k})}=\left(1-\ket{u_+(\textbf{k})}\bra{u_+(\textbf{k})}\right)\ket{\partial_k u(k)}\\
    &=\frac{\sqrt{2}(f_1'(k)f_2(k)-f_1(k)f_2'(k))}{(|f_1(k)|^2+|f_2(k)|^2)^2}
    \ket{z(k)},
\end{split}
\end{align}
where $\ket{z(k)}$ is given in Eq.~(\ref{eq:generalthreep}).
For Euler curvature to be non zero for all $k$, $\forall k \bra{v(\textbf{k})}\ket{v(\textbf{k})}\neq 0$. If $f_1(k)f_2'(k)-f_1'(k)f_2(k)$ is not a constant, there exists a complex number $k$ that satisfies $f_1(k)f_2'(k)-f_1'(k)f_2(k)=0$ i.e.  $\bra{v(\textbf{k})}\ket{v(\textbf{k})}=0$. Therefore
\begin{align}
\begin{split}
    &(f_1(k)f_2'(k)-f_1'(k)f_2(k))'\\
    &=f_1(k)f_2''(k)-f_1''(k)f_2(k)=0.
\end{split}
\end{align}
For $f_1(k)f_2''(k)=f_1''(k)f_2(k)$ to hold while $f_1(k),f_2(k)$ do not share roots, $f_1''(k)$ should be zero for k which $f_1(k)$ is zero. It holds only when there exists a polynomial $g(k)$ such that $f_1''(k)=f_1(k)g(k)$. Since the degree of $f_1''(k)$ is lower than $f_1(k)$, it is only possible if $f_1''(k)=0$. It is the same for $f_2(k)$. Therefore, the degree of $f_1(k),f_2(k)$ should be lower than two. Therefore, $f_1(k),f_2(k)$ becomes linear expressions of k while $f_1(k),f_2(k)$ do not share roots. By Eq. (\ref{eq:param1}), Eq. (\ref{eq:param2}), Eq. (\ref{eq:param3}), when $f_1(k),f_2(k)$ are linear for $k$, the highest degree of $\tilde{u}(k)_1,\tilde{u}(k)_2,\tilde{u}(k)_3$ is 2. As a result, the Euler number of the ideal Euler band in the three-band model should be $\pm2$. 

\begin{figure}
	\includegraphics[width=0.47\textwidth]{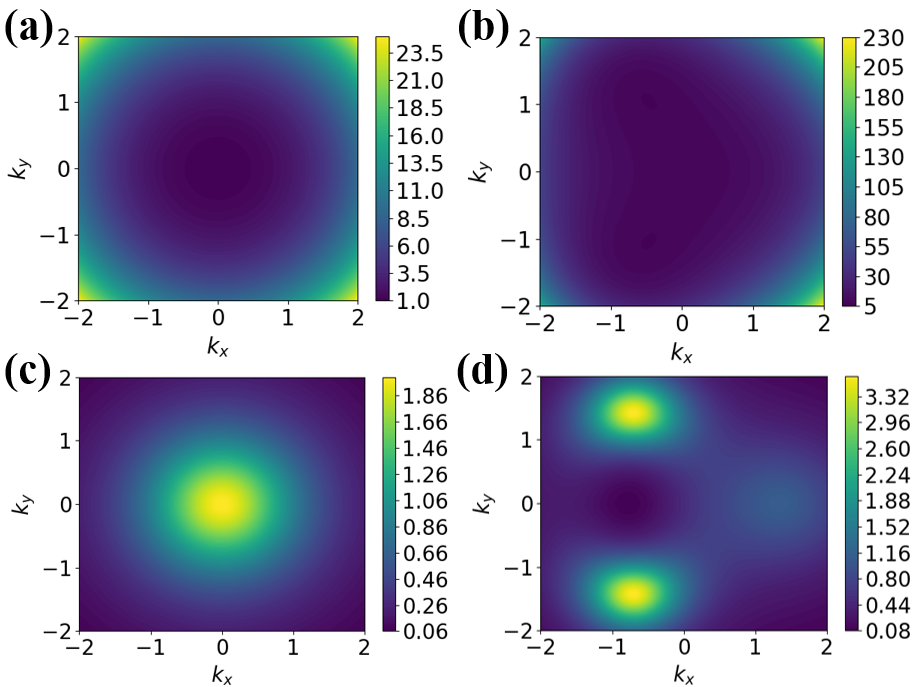}
  \vspace{-0.3cm}
	\caption{(a) Plot of the bandgap between second and third band for the three band model given in Eq.~(\ref{eq:threeBandH}) (b) Plot of the bandgap between second and third band for the four band model given in Eq.~(\ref{eq:fourBandH}) with $n=3$. For both cases, the band gap is opened. These gaps describe the whole band structure because, the lower two bands for these models are flat and degenerate while having energy eigenvalue zero, and the upper two bands for the four-band model is degenerate. (c) Plot of the Euler curvature of lower two bands in the three band model given in Eq.~(\ref{eq:threeBandH}) (d) Same plot for the four band model given in (\ref{eq:fourBandH}) with $n=3$.}
 \label{fig:Ideal}
\end{figure}

One example of the three-band Hamiltonian having the ideal Euler band with Euler number $\pm2$ as lower two bands can be constructed by,
\begin{align}
    f_1(k)=k, f_2(k)=1,
\end{align}
and Eq. (\ref{eq:generalthree}), Eq. (\ref{eq:generalthreep}).
\begin{align}
\label{eq:threeBandH}
    &H(\textbf{k})=\ket{z(k)}\bra{z(k)},\\
    &\ket{z(k)}=\begin{pmatrix}
        k+k^*\\
        \frac{1}{i}(k^*-k)\\
        1-|k|^2
    \end{pmatrix}.
\end{align}
Fig.~\ref{fig:Ideal}(b) shows the band gap and Fig.~\ref{fig:Ideal}(c) shows the Euler curvature for lower two bands. Both values do not become zero.

If the Euler curvature is allowed to be 0, the ideal Euler band in the three band model can have any even number as the Euler number. We can show this by example
\begin{align}
    f_1(k)=k^n, f_2(k)=1.
\end{align}
Applying this to Eq. (\ref{eq:generalthree}) and Eq. (\ref{eq:generalthreep}) gives Hamiltonian
\begin{align}
\label{eq:threeBandH2}
    &H(\textbf{k})=\ket{z(k)}\bra{z(k)},\\
    &\ket{z(k)}=\begin{pmatrix}
        k^n+(k^*)^n\\
        \frac{1}{i}((k^*)^n-k^n)\\
        1-|k|^{2n}
    \end{pmatrix}.
\end{align}
Since the highest degree of $\tilde{u}(k)_1,\tilde{u}(k)_2,\tilde{u}(k)_3$ is $2n$, the lower two bands of this Hamiltonian is the ideal Euler band with Euler number $\pm 2n$.

\subsection{Four band model}
For the ideal Euler bands in the three-band model, the Euler number had to be $\pm2$. Here we construct a four-band model having the ideal Euler bands with an arbitrary Euler number. The ideal Euler band $\ket{u_1(\textbf{k})}$, $\ket{u_2(\textbf{k})}$ in four band model can be represented as,
\begin{align}
    &\frac{1}{\sqrt{2}}\left(\ket{u_1(\textbf{k})}+i\ket{u_2(\textbf{k})}\right)=\frac{1}{N_{\textbf{k}}}\ket{\tilde{u}(k)}=\frac{1}{N_{\textbf{k}}}
    \begin{pmatrix}
        \tilde{u}(k)_1\\
        \tilde{u}(k)_2\\
        \tilde{u}(k)_3\\
        \tilde{u}(k)_4
    \end{pmatrix},
\end{align}
where $\tilde{u}(k)_1$, $\tilde{u}(k)_2$, $\tilde{u}(k)_3$, $\tilde{u}(k)_4$, are holomorphic functions for $k$.
From Eq.~(\ref{eq:condition3}), 
\begin{align}
\label{eq:conditionFourband}
\begin{split}
    &(\tilde{u}(k)_1+i\tilde{u}(k)_2)(\tilde{u}(k)_1-i\tilde{u}(k)_2)\\
    &=-(\tilde{u}(k)_3+i\tilde{u}(k)_4)(\tilde{u}(k)_3-i\tilde{u}(k)_4)
\end{split}
\end{align}
By using polynomials $g_1(k), g_2(k), g_3(k), g_4(k)$ defined as
\begin{align}
    &\tilde{u}(k)_1=g_1(k)+g_2(k),\\
    &\tilde{u}(k)_2=i(g_1(k)-g_2(k)),\\
    &\tilde{u}(k)_3=g_3(k)-g_4(k),\\
    &\tilde{u}(k)_4=i(g_3(k)+g_4(k)),
\end{align}
the condition in Eq.~(\ref{eq:conditionFourband}) can be simplified as
\begin{align}
    &g_1(k)g_2(k)=g_3(k)g_4(k).
\end{align}

For the four band model, the Euler number for $\ket{u_1(\textbf{k})}$, $\ket{u_2(\textbf{k})}$ can be arbitrary integer. We prove this by constructing an example that has an arbitrary integer as the Euler number. The Euler number for $\ket{u_1(\textbf{k})},\ket{u_2(\textbf{k})}$ is the highest order of $k$ among $\tilde{u}(k)_1,\tilde{u}(k)_2,\tilde{u}(k)_3,\tilde{u}(k)_4$.

Therefore for
\begin{align}
\label{eq:sample1}
    &g_1(k)=k^n-1,\\
\label{eq:sample2}
    &g_2(k)=-1,\\
\label{eq:sample3}
    &g_3(k)=\frac{k^n-1}{k-1},\\
\label{eq:sample4}
    &g_4(k)=-k+1.
\end{align}
the Euler number is $\pm n$. Eq.~(\ref{eq:condition2}) holds because
\begin{align}
\begin{split}
\bra{\tilde{u}(k)}\ket{\tilde{u}(k)}&=\sum_{i=1}^4|\tilde{u}_i(k)|^2=2\sum_{i=1}^4|\tilde{g}_i(k)|^2\\
&\geq 2|g_2(k)|^2=1>0.
\end{split}
\end{align}

The Hamiltonian having $\ket{u_1(\textbf{k})}, \ket{u_2(\textbf{k})}$ as lower two bands can be constructed by
\begin{align}
\label{eq:fourBandH}
    &H(\textbf{k})=\ket{\tilde{w}(k,k^*)}\bra{\tilde{w}(k,k^*)}+\left(\ket{\tilde{w}(k,k^*)}\bra{\tilde{w}(k,k^*)}\right)^*,\\
    \label{eq:wvector}
    &\ket{\tilde{w}(k,k^*)}=\frac{k^n-1}{k-1}
    \begin{pmatrix}
    1\\
    i\\
    -k^*+1\\
    -ik^*+i
    \end{pmatrix}+
    \begin{pmatrix}
    k^*-1\\
    -ik^*+i\\
    1\\
    -i
    \end{pmatrix}.
\end{align}
The Hamiltonian is real. Since $\ket{\tilde{w}(k,k^*)}$ is orthogonal to both $\ket{u_1(\textbf{k})},\ket{u_2(\textbf{k})}$, the energy eigenvalues for the lower two band is zero and the eigenstates for the lower two bands are $\ket{u_1(\textbf{k})},\ket{u_2(\textbf{k})}$. The upper two bands are degenerate with the energy eigenvalue,
\begin{align}
    \bra{\tilde{w}(k,k^*)}\ket{\tilde{w}(k,k^*)}=2(1+|k-1|^2)\left(1+\left|\frac{k^n-1}{k-1}\right|^2\right)>0.
\end{align}
Therefore, the band gap is opened. Fig.~\ref{fig:Ideal}(b) shows the band gap when $n=3$ and Fig.~\ref{fig:Ideal}(d) shows the Euler curvature for the lower two bands when $n=3$. Both values do not become zero.

Condition from Eq.~(\ref{eq:curvatureCondition}), $\forall k,  F_{12}(\textbf{k})\neq 0$ also holds. The proof is as follows. Since $\ket{\tilde{w}(k,k^*)}$ is orthogonal to both $\ket{u_1(\textbf{k})},\ket{u_2(\textbf{k})}$, $\bra{u_+(\textbf{k})}\ket{\tilde{w}(k,k^*)}$ is zero. Therefore, from Eq.~(\ref{eq:HolomorphicEuler}), Eq.~(\ref{eq:vforvendetta}),
\begin{align}
\begin{split}
    F_{12}(\textbf{k})&=\frac{1}{N_\textbf{k}^2}\bra{\partial_k \tilde{u}(k)}(1-\ket{u_+(\textbf{k})}\bra{u_+(\textbf{k})})\ket{\partial_k \tilde{u}(k)}\\
    &\geq \frac{1}{N_\textbf{k}^2}\frac{\bra{\partial_k \tilde{u}(k)}\ket{\tilde{w}(k,k^*)}\bra{\tilde{w}(k,k^*)}\ket{\partial_k \tilde{u}(k)}}{\bra{\tilde{w}(k,k^*)}\ket{\tilde{w}(k,k^*)}}.
\end{split}
\end{align}
If we calculate $\bra{\tilde{w}(k,k^*)}\ket{\partial_k u(k)}$ using Eq. (\ref{eq:sample1}), Eq.~(\ref{eq:sample2}), Eq.~(\ref{eq:sample3}), Eq.~(\ref{eq:sample4}), Eq.~(\ref{eq:wvector}), it becomes
\begin{align}
\begin{split}
    &\bra{\tilde{w}(k,k^*)}\ket{\partial_k u(k)}=2\left(1+\left|\frac{k^n-1}{k-1}\right|^2\right)>0.
\end{split}
\end{align}
Therefore $\forall k, F_{12}(\textbf{k})> 0$. 

The upper two bands also have the Euler number. Unlike the Chern number the sum of the Euler number does not always become zero, therefore it should be calculated separately. For the upper two bands similar equation with Eq.~(\ref{eq:ChernBasis}) holds.
\begin{align}
    \frac{1}{\sqrt{2}}\left(\ket{u_3(\textbf{k})}+i\ket{u_4(\textbf{k})}\right)=\frac{1}{N_{\textbf{k}}^w}\ket{\tilde{w}(k,k^*)},
\end{align}
where $N_\textbf{k}^w$ is a normalization factor. Since $\ket{\tilde{w}(k,k^*)}$ is not a holomorphic function of k, we can't use Eq.~(\ref{eq:e2k}) directly to calculate the Euler number. Instead by following the same procedure used to derive Eq.~(\ref{eq:e2k}), 

\begin{align}
\label{eq:eulerkkstar}
\begin{split}
    e_2&=\frac{1}{2\pi i}\oint \frac{\bra{\tilde{w}(k,k^*)}\ket{\partial_k\tilde{w}(k,k^*)}}{\bra{\tilde{w}(k,k^*)}\ket{\tilde{w}(k,k^*)}}dk\\
    &+\frac{1}{2\pi i}\oint \frac{\bra{\tilde{w}(k,k^*)}\ket{\partial_{k^*}\tilde{w}(k,k^*)}}{\bra{\tilde{w}(k,k^*)}\ket{\tilde{w}(k,k^*)}}dk^*
\end{split}
\end{align}
can be derived, where the contour integral is done along a circle whose radius goes to infinity. The direction of the contour integral for $k$ and $k^*$ is opposite. From Eq.~(\ref{eq:wvector}), at the boundary of the complex plane
\begin{align}
\label{eq:largek4}
    \bra{\tilde{w}(k,k^*)}\ket{\partial_k\tilde{w}(k,k^*)}\rightarrow 2|k|^{2n}\frac{n-1}{k},\\
\label{eq:largek44}
    \bra{\tilde{w}(k,k^*)}\ket{\partial_k^*\tilde{w}(k,k^*)}\rightarrow 2|k|^{2n}\frac{1}{k^*},\\
\label{eq:largek444}
    \bra{\tilde{w}(k,k^*)}\ket{\tilde{w}(k,k^*)}\rightarrow 2|k|^{2n}.
\end{align}
From Eq.~(\ref{eq:eulerkkstar}), Eq.~(\ref{eq:largek4}),  Eq.~(\ref{eq:largek44}), Eq.~(\ref{eq:largek444}), the Euler number for upper two band is $\pm(n-2)$.

\section{Idealness and GMP algebra}
Due to the interaction, the ground state with a fractional Chern number is expected to have a gap, even at fractional filling. Therefore, determining whether there is a many-body gap is important when constructing a model with fractional topology. In the case of fractionally filled Lowest Landau level (LLL), GMP algebra predicts the dispersion well enough to determine whether the ground state with Laughlin wavefunction is gapped i.e. stable~\cite{GMPLLL}. If the lattice model has the same algebra, for the nearly flat bands it is likely to have a many-body gap at fractional filling equal to the filling factor that LLL has a many-body gap. For the lattice model, it was shown that the GMP algebra holds if a Chern band is ideal and has uniform Berry curvature~\cite{GMPRoy}. In this section, we first review the relation between the idealness and GMP algebra for the Chern insulator and prove that a similar relation exists for the Chern basis of the Euler bands.

\subsection{Quantum geometry from position operators}
Before diving into the GMP algebra, it is helpful to obtain an equation for a quantum geometric tensor composed of position operators. It starts by writing the Bloch wave function using the real-space basis. For the N band tight-binding model, the Bloch wavefunction is written as 
\begin{align}
\ket{u_n(\textbf{k})}=\begin{pmatrix}
    u_n(\textbf{k})_1\\
    u_n(\textbf{k})_2\\
    .\\
    .\\
    .\\
    u_n(\textbf{k})_N\\
\end{pmatrix}.
\end{align}
This wavefunction can be written in orbital basis as,
\begin{align}
   \label{eq:tightBinding} &\ket{\psi_n(\textbf{k})}=\sum_{\alpha}u_n(\textbf{k})_\alpha\ket{\textbf{k},\alpha},\\
    &\ket{\textbf{k},\alpha}=\frac{A_{unit}}{(2\pi)^2}\sum_{\textbf{R}}e^{i\textbf{k}\cdot(\textbf{R}+\textbf{x}_\alpha)}\ket{\textbf{R},\alpha},
\end{align}
where $\textbf{R}$ is a Bravais lattice vector, $\textbf{x}_\alpha$ is a position of the orbital $\alpha$ in the unit cell, and $A_{unit}$ is the area of the unit cell. The wavefunction is normalized by,
\begin{align}
    \bra{\textbf{k}',\alpha}\ket{\textbf{k},\beta}=\delta^2(\textbf{k}'-\textbf{k})\delta_{\alpha\beta},\\
    \bra{\psi_n(\textbf{k}')}\ket{\psi_m(\textbf{k})}=\delta^2(\textbf{k}'-\textbf{k})\delta_{nm}.
\end{align}
The position operator is applied as,
\begin{align}
\label{eq:position}
    \begin{split}
        \textbf{r}_\mu\ket{\psi_n(\textbf{k})}&=\frac{A_{unit}}{(2\pi)^2}\sum_{\alpha \textbf{R}}(\textbf{R}+\textbf{x}_\alpha)_be^{i\textbf{k}\cdot(\textbf{R}+\textbf{x}_\alpha)}u_n(\textbf{k})_\alpha\ket{\textbf{R},\alpha}\\
        &=\frac{1}{i}\partial_{\textbf{k}_\mu}\ket{\psi_n(\textbf{k})}-\frac{1}{i}\sum_{\alpha}(\partial_{\textbf{k}_\mu}u_n(\textbf{k})_\alpha)\ket{\textbf{k},\alpha}.\\
    \end{split}
\end{align}
From Eq.~(\ref{eq:position})
\begin{align}
\label{eq:position2}
    \begin{split}
    &\bra{\psi_n(\textbf{k}')}\textbf{r}_\mu\textbf{r}_\nu\ket{\psi_n(\textbf{k})}\\
    &=\left(\partial_{\textbf{k}_\mu'}\bra{\psi_n(\textbf{k}')}-\sum_{\alpha}(\partial_{\textbf{k}_\mu'}u_n^*(\textbf{k}')_\alpha)\bra{\textbf{k}',\alpha}\right)\\
    &\left(\partial_{\textbf{k}_\nu}\ket{\psi_n(\textbf{k})}-\sum_{\alpha}(\partial_{\textbf{k}_\nu}u_n(\textbf{k})_\alpha)\ket{\textbf{k},\alpha}\right)\\
    &=\partial_{\textbf{k}_\mu'}\partial_{\textbf{k}_\nu}\bra{\psi_n(\textbf{k}')}\ket{\psi_n(\textbf{k})}\\
    &-\sum_{\alpha}(\partial_{\textbf{k}_\mu'}u_n^*(\textbf{k}')_\alpha)(\partial_{\textbf{k}_\nu}\bra{\textbf{k}',\alpha}\ket{\psi_n(\textbf{k})})\\
    &-\sum_{\alpha}(\partial_{\textbf{k}_\nu}u_n(\textbf{k})_\alpha)(\partial_{\textbf{k}_\mu'}\bra{\psi_n(\textbf{k}')}\ket{\textbf{k},\alpha})\\
    &+\bra{\partial_{\textbf{k}_\mu}u_n(\textbf{k})}\ket{\partial_{\textbf{k}_\nu}u_n(\textbf{k})}\delta^2(\textbf{k}'-\textbf{k}).
    \end{split}
\end{align}
Also, by defining the projection operator for the Chern bands as
\begin{align}
    P_n=\int \ket{\psi_n(\textbf{k})}\bra{\psi_n(\textbf{k})}d\textbf{k},
\end{align}
\begin{align}
\label{eq:position3}
    \begin{split}
    &\bra{\psi_n(\textbf{k}')}\textbf{r}_\mu P_n\textbf{r}_\nu\ket{\psi_n(\textbf{k})}=\\
    &\int\left(\partial_{\textbf{k}_\mu'}\bra{\psi_n(\textbf{k}')}\ket{\psi_n(\textbf{k}'')}-\sum_{\alpha}(\partial_{\textbf{k}_\mu'}u_n^*(\textbf{k}')_\alpha)\bra{\textbf{k}',\alpha}\ket{\psi_n(\textbf{k}'')}\right)\\
    &\left(\partial_{\textbf{k}_\nu}\bra{\psi_n(\textbf{k}'')}\ket{\psi_n(\textbf{k})}-\sum_{\alpha}(\partial_{\textbf{k}_\nu}u_n(\textbf{k})_\alpha)\bra{\psi_n(\textbf{k}'')}\ket{\textbf{k},\alpha}\right)d\textbf{k}''\\
    &=\int d\textbf{k}''(\partial_{\textbf{k}_\mu'}\delta^2(\textbf{k}'-\textbf{k}''))(\partial_{\textbf{k}_\nu}\delta^2(\textbf{k}''-\textbf{k}))\\
    &-\sum_{\alpha}(\partial_{\textbf{k}_\mu'}u_n^*(\textbf{k}')_\alpha)(\bra{\textbf{k}',\alpha}\ket{\psi_n(\textbf{k}'')})\partial_{\textbf{k}_\nu}\delta^2(\textbf{k}''-\textbf{k})\\
    &-\sum_{\alpha}(\partial_{\textbf{k}_\nu}u_n(\textbf{k}')_\alpha)(\bra{\psi_n(\textbf{k}'')}\ket{\textbf{k},\alpha})\partial_{\textbf{k}_\mu'}\delta^2(\textbf{k}''-\textbf{k}')\\
    &+\sum_{\alpha\beta}(\partial_{\textbf{k}_\mu'}u_n^*(\textbf{k}')_\alpha)(\partial_cu_n(\textbf{k})_\beta)\bra{\textbf{k}',\alpha}\ket{\psi_n(\textbf{k}'')}\bra{\psi_n(\textbf{k}'')}\ket{\textbf{k},\beta}\\
    &=\int -(\partial_{\textbf{k}_\mu''}\delta^2(\textbf{k}'-\textbf{k}''))(\partial_{\textbf{k}_\nu}\delta^2(\textbf{k}''-\textbf{k}))\\
    &+\sum_{\alpha}(\partial_{\textbf{k}_\mu'}u_n^*(\textbf{k}')_\alpha)(\bra{\textbf{k}',\alpha}\ket{\psi_n(\textbf{k}'')})\partial_{\textbf{k}_\nu''}\delta^2(\textbf{k}''-\textbf{k})\\
    &+\sum_{\alpha}(\partial_{\textbf{k}_\nu}u_n(\textbf{k}')_\alpha)(\bra{\psi_n(\textbf{k}'')}\ket{\textbf{k},\alpha})\partial_{\textbf{k}_\mu''}\delta^2(\textbf{k}''-\textbf{k}')\\
    &+\bra{\partial_{\textbf{k}_\mu'}u_n^*(\textbf{k}')}\ket{u_n(\textbf{k}'')}\bra{u_n(\textbf{k}'')}\ket{\partial_{\textbf{k}_\nu}u_n(\textbf{k})}\\
    &\qquad \delta^2(\textbf{k}'-\textbf{k}'')\delta^2(\textbf{k}''-\textbf{k})d\textbf{k}''\\
    &=\partial_{\textbf{k}_\mu'}\partial_{\textbf{k}_\nu}\delta^2(\textbf{k}'-\textbf{k})\\
    &-\sum_{\alpha}(\partial_{\textbf{k}_\mu'}u_n^*(\textbf{k}')_\alpha)(\partial_{\textbf{k}_\nu}\bra{\textbf{k}',\alpha}\ket{\psi_n(\textbf{k})})\\
    &-\sum_{\alpha}(\partial_{\textbf{k}_\nu}u_n(\textbf{k})_\alpha)(\partial_{\textbf{k}_\mu'}\bra{\psi_n(\textbf{k}')}\ket{\textbf{k},\alpha})\\
    &+\bra{\partial_{\textbf{k}_\mu'}u_n^*(\textbf{k}')}\ket{u_n(\textbf{k}'')}\bra{u_n(\textbf{k}'')}\ket{\partial_{\textbf{k}_\nu}u_n(\textbf{k})}\delta^2(\textbf{k}'-\textbf{k}).\\
    \end{split}
\end{align}

Subtracting Eq.~(\ref{eq:position3}) from Eq.~(\ref{eq:position2}) gives the expression of the quantum geometric tensor composed of position operators.
\begin{align}
\label{eq:positionGeometry}
\begin{split}
\bra{\psi_n(\textbf{k}')}\textbf{r}_\mu(I-P_n)\textbf{r}_\nu\ket{\psi_n(\textbf{k})}=Q_{\mu\nu}(\textbf{k})\delta^2(\textbf{k}'-\textbf{k}).
\end{split}
\end{align}

\subsection{GMP algebra and the ideal Chern band}
If the band is flat, the Hamiltonian for electrons only has the interacting part. In this case, the whole Hamiltonian can be written by the density operators. Additionally, if the band gap is larger than the interaction, mixing between occupied and unoccupied bands can be neglected. In this case, instead of the conventional density operator $\rho(\textbf{k})=e^{i\textbf{k}\cdot\textbf{r}}$, the density operator projected to the occupied band can be used to describe the interaction. Therefore, the commutation relation between the projected density operators is important for determining the energy spectrum. The commutation relation that is used to determine the energy spectrum of LLL with fractional filling is called the GMP algebra~\cite{GMPRoy}. The GMP algebra is given as,
\begin{align}
\begin{split}
&[\bar{\rho}_n(\textbf{k}),\bar{\rho}_n(\textbf{q})]\\
&=2ie^{\textbf{k}_\mu g_{\mu\nu}^n\textbf{q}_\nu}\sin\left(\left(\textbf{k}_x\textbf{q}_y-\textbf{k}_y\textbf{q}_x\right)\Omega_{xy}^n\right)\bar{\rho}_n(\textbf{k}+\textbf{q}),
\end{split}
\end{align}
where the projected density operator $\bar{\rho}_n(\textbf{k})=P_ne^{i\textbf{k}\cdot\textbf{r}}P_n$. The Chern band which satisfies this algebra is expected to have a many-body gap at fractional filling as in the LLL.
 
If the Berry curvature is uniform and the Chern band is ideal, there exists the coordinate in which the GMP algebra holds. The coordinate in which the GMP algebra holds can be found by
\begin{align}
\label{eq:newcoordinate1}
    &\textbf{r}'_{\mu}=\sum_\nu t_{\mu\nu} \textbf{r}_\nu,\\
\label{eq:newcoordinate2}
    &\frac{\Omega}{|\Omega|}\delta_{\mu\nu}=\sum_{\lambda\rho}t_{\mu\lambda}\omega_{\lambda\rho}t_{\nu\rho},\\
    &g_{\mu\nu}=\frac{1}{2}\Omega\omega_{\mu\nu},
\end{align}
where $\textbf{r}'_a$ is a new coordinate and $t_{\mu\nu}$ is a real and invertible matrix. Since the determinant of $\omega_{\mu\nu}$ is one, $t_{\mu\nu}$ exists.

When the Berry curvature is positive, under this new coordinate, the quantum geometric tensor becomes
\begin{align}
\label{eq:newcoordinate}
    Q_{\mu\nu}(\textbf{k})=\frac{1}{2}\Omega_{xy}^n(\textbf{k})
    \begin{pmatrix}
        1 & -i\\
        i & 1\\
    \end{pmatrix}.
\end{align}
For the new coordinate, from Eq.~(\ref{eq:positionGeometry}) and Eq.~(\ref{eq:newcoordinate}),
\begin{align}
\label{eq:zero}
\begin{split}
&\bra{\psi_n(\textbf{k}')}(x+iy)(I-P_n)(x-iy)\ket{\psi_n(\textbf{k})}\\
&=\frac{1}{2}\Omega_{xy}^n(\textbf{k})\delta^2(\textbf{k}-\textbf{k}')
\begin{pmatrix}
    1 & i\\
\end{pmatrix}
    \begin{pmatrix}
        1 & -i\\
        i & 1\\
    \end{pmatrix}\begin{pmatrix}
    1\\
    -i\\
\end{pmatrix}=0.
\end{split}
\end{align}
This implies $D^{\dagger}D=0$ i.e. $D=0$, where $D=(I-P_n)(x-iy)P_n$. Proving that GMP algebra holds from these results is straightforward. From Eq.~(\ref{eq:zero}),
\begin{align}
\begin{split}
&\textbf{r}_-P_n=P_n\textbf{r}_-P_n,\\
&P_n\textbf{r}_+=\left(\textbf{r}_-P_n\right)^{\dagger}=\left(P_n\textbf{r}_-P_n\right)^{\dagger}=P_n\textbf{r}_+P_n,\\
\end{split}
\end{align}
 where $\textbf{r}_{\pm}=x\pm iy$. Using this relation and $[\textbf{r}_+,\textbf{r}_-]=0$,
\begin{align}
\begin{split}\label{eq:density}
\bar{\rho}_n(\textbf{k})&=P_ne^{i\textbf{k}\cdot\textbf{r}}P_n=P_ne^{i\textbf{k}_-\textbf{r}_+/2+i\textbf{k}_+\textbf{r}_-/2}P_n\\
&=P_ne^{i\textbf{k}_-\textbf{r}_+/2}e^{i\textbf{k}_+\textbf{r}_-/2}P_n\\
&=P_ne^{i\textbf{k}_-\textbf{r}_+/2}P_ne^{i\textbf{k}_+\textbf{r}_-/2}P_n\\
&=e^{i\textbf{k}_-P_n\textbf{r}_+P_n/2} e^{i\textbf{k}_+P_n\textbf{r}_-P_n/2},
\end{split}
\end{align}
where $\textbf{k}_{\pm}=\textbf{k}_x\pm i\textbf{k}_y$. Now the commutator between density operators can be calculated from the commutator between $P_n\textbf{r}_+P_n$ and $P_n\textbf{r}_-P_n$. This commutator can be obtained from Eq.~(\ref{eq:positionGeometry}) as, 
\begin{align}
\label{eq:positionCommute}
[P_n\textbf{r}_+P_n,P_n\textbf{r}_-P_n]=2\Omega^n_{xy}P_n. 
\end{align}
From Eq.~(\ref{eq:density}) and Eq.~(\ref{eq:positionCommute}), the commutator between density operators becomes
\begin{align}
\begin{split}
&[\bar{\rho}_n(\textbf{k}),\bar{\rho}_n(\textbf{q})]\\
&=2ie^{\textbf{k}_\mu g_{\mu\nu}^n\textbf{q}_\nu}\sin\left(\left(\textbf{k}_x\textbf{q}_y-\textbf{k}_y\textbf{q}_x\right)\Omega_{xy}^n/2\right)\bar{\rho}_n(\textbf{k}+\textbf{q}).
\end{split}
\end{align}

When the Berry curvature is negative, the quantum geometric tensor becomes
\begin{align}
\label{eq:newcoordinateQ2}
    Q_{\mu\nu}(\textbf{k})=\frac{1}{2}\Omega_{xy}^n(\textbf{k})
    \begin{pmatrix}
        -1 & -i\\
        i & -1\\
    \end{pmatrix},
\end{align}
under the new coordinate. For the new coordinate, from Eq.~(\ref{eq:positionGeometry}) and Eq.~(\ref{eq:newcoordinateQ2}),
\begin{align}
\label{eq:zero2}
\begin{split}
&\bra{\psi_n(\textbf{k}')}(x-iy)(I-P_n)(x+iy)\ket{\psi_n(\textbf{k})}\\
&=\frac{1}{2}\Omega_{xy}^n(\textbf{k})\delta^2(\textbf{k}-\textbf{k}')
\begin{pmatrix}
    1 & -i\\
\end{pmatrix}
    \begin{pmatrix}
        -1 & -i\\
        i & -1\\
    \end{pmatrix}\begin{pmatrix}
    1\\
    i\\
\end{pmatrix}=0.
\end{split}
\end{align}
From Eq.~(\ref{eq:zero2}),
\begin{align}
\begin{split}
&\textbf{r}_+P_n=P_n\textbf{r}_+P_n,\\
&P_n\textbf{r}_-=\left(\textbf{r}_+P_n\right)^{\dagger}=\left(P_n\textbf{r}_+P_n\right)^{\dagger}=P_n\textbf{r}_-P_n,\\
\end{split}
\end{align}
 where $\textbf{r}_{\pm}=x\pm iy$. Using this relation and $[\textbf{r}_+,\textbf{r}_-]=0$,
\begin{align}
\begin{split}\label{eq:density2}
\bar{\rho}_n(\textbf{k})&=P_ne^{i\textbf{k}\cdot\textbf{r}}P_n=P_ne^{i\textbf{k}_+\textbf{r}_-/2+i\textbf{k}_-\textbf{r}_+/2}P_n\\
&=P_ne^{i\textbf{k}_+\textbf{r}_-/2}e^{i\textbf{k}_-\textbf{r}_+/2}P_n\\
&=P_ne^{i\textbf{k}_+\textbf{r}_-/2}P_ne^{i\textbf{k}_-\textbf{r}_+/2}P_n\\
&=e^{i\textbf{k}_+P_n\textbf{r}_-P_n/2} e^{i\textbf{k}_-P_n\textbf{r}_+P_n/2},
\end{split}
\end{align}
 where $\textbf{k}_\pm=\textbf{k}_x\pm i\textbf{k}_y$. As before, from Eq. (\ref{eq:positionCommute}), Eq. (\ref{eq:density2}),
\begin{align}
\begin{split}
&[\bar{\rho}_n(\textbf{k}),\bar{\rho}_n(\textbf{q})]\\
&=2ie^{\textbf{k}_\mu g_{\mu\nu}^n\textbf{q}_\nu}\sin\left(\left(\textbf{k}_x\textbf{q}_y-\textbf{k}_y\textbf{q}_x\right)\Omega_{xy}^n(\textbf{k})/2\right)\bar{\rho}_n(\textbf{k}+\textbf{q}).
\end{split}
\end{align}
 
\subsection{GMP algebra and the ideal Euler band}
In this section, we show that when the Euler curvature is uniform for the ideal Euler bands, the coordinate in which the GMP algebra holds exists for the density operator projected to the Chern basis. If the Euler curvature is uniform and unit determinant $\omega_{\mu\nu}$ s.t. $g_{\mu\nu}(\textbf{k})=F_{12}(\textbf{k})\omega_{\mu\nu}$ exists for the isolated two bands then,
\begin{align}
\label{eq:connectingIdeal}
    &g^+_{\mu\nu}(\textbf{k})=\frac{1}{2}g_{\mu\nu}(\textbf{k})=\frac{1}{2}F_{12}(\textbf{k})\omega_{\mu\nu}=\frac{1}{2}\Omega^+_{xy}(\textbf{k})(-\omega_{\mu\nu}),\\
\label{eq:connectingIdeal2}
    &g^-_{\mu\nu}(\textbf{k})=\frac{1}{2}g_{\mu\nu}(\textbf{k})=\frac{1}{2}F_{12}(\textbf{k})\omega_{\mu\nu}=\frac{1}{2}\Omega^-_{xy}(\textbf{k})\omega_{\mu\nu},
\end{align}
and $\Omega^\pm_{xy}(\textbf{k})$ is uniform by Eq.~(\ref{eq:connectingCurvature}), Eq.~(\ref{eq:connectingMetric}). From Eq. (\ref{eq:newcoordinate1}), Eq. (\ref{eq:newcoordinate2}), the coordinate in which the GMP algebra holds for each Chern basis is the same. In this coordinate,
\begin{align}
\label{eq:pGMP}
\begin{split}
&[\rho_{++}(\textbf{k}),\rho_{++}(\textbf{q})]\\
&=2ie^{\textbf{k}_\mu g_{\mu\nu}^+\textbf{q}_\nu}\sin\left(\left(\textbf{k}_x\textbf{q}_y-\textbf{k}_y\textbf{q}_x\right)\Omega_{xy}^+/2\right)\rho_{++}(\textbf{k}+\textbf{q}),
\end{split}\\
\label{eq:mGMP}
\begin{split}
&[\rho_{--}(\textbf{k}),\rho_{--}(\textbf{q})]\\
&=2ie^{\textbf{k}_\mu g_{\mu\nu}^-\textbf{q}_\nu}\sin\left(\left(\textbf{k}_x\textbf{q}_y-\textbf{k}_y\textbf{q}_x\right)\Omega_{xy}^-/2\right)\rho_{--}(\textbf{k}+\textbf{q}),
\end{split}
\end{align}
where
\begin{align}
&\rho_{\alpha\beta}(\textbf{k})=P_\alpha e^{i\textbf{k}\cdot\textbf{r}} P_\beta,\\
&P_{\pm}=\int\ket{\psi_\pm(\textbf{k})}\bra{\psi_\pm(\textbf{k})}d\textbf{k},\\
&\ket{\psi_\pm(\textbf{k})}=e^{i\textbf{k}\cdot\textbf{r}}(\ket{u_1(\textbf{k})}\pm i\ket{u_2(\textbf{k})})/\sqrt{2}
\end{align}
Therefore the density operator projected to each Chern basis has GMP algebra. But the total density operator for the two bands
\begin{align}
\label{eq:totaldensity}
&\bar{\rho}(\textbf{k})=\rho_{++}(\textbf{k})+\rho_{+-}(\textbf{k})+\rho_{-+}(\textbf{k})+\rho_{--}(\textbf{k}),
\end{align}
has $\rho_{-+}(\textbf{k})$, $\rho_{+-}(\textbf{k})$. If the sign of the Berry curvature for the two Chern bases are both positive,
\begin{align}
\begin{split}
\rho_{+-}(\textbf{k})&=P_+e^{i\textbf{k}\cdot\textbf{r}}P_-=P_+e^{i\textbf{k}_-\textbf{r}_+/2+i\textbf{k}_+\textbf{r}_-/2}P_-\\
&=P_+e^{i\textbf{k}_-\textbf{r}_+/2}e^{i\textbf{k}_+\textbf{r}_-/2}P_-\\
&=P_+e^{i\textbf{k}_-\textbf{r}_+/2}P_+P_-e^{i\textbf{k}_+\textbf{r}_-/2}P_-\\
&=0.
\end{split}\\
\begin{split}
\rho_{-+}(\textbf{k})&=P_-e^{i\textbf{k}\cdot\textbf{r}}P_+=P_-e^{i\textbf{k}_-\textbf{r}_+/2+i\textbf{k}_+\textbf{r}_-/2}P_+\\
&=P_-e^{i\textbf{k}_-\textbf{r}_+/2}e^{i\textbf{k}_+\textbf{r}_-/2}P_+\\
&=P_-e^{i\textbf{k}_-\textbf{r}_+/2}P_-P_+e^{i\textbf{k}_+\textbf{r}_-/2}P_+\\
&=0.
\end{split}
\end{align}
The same applies when both signs are negative. However, because the signs are opposite, generally $\rho_{-+}(\textbf{k})$, $\rho_{+-}(\textbf{k})$ are not zero. This prevents the commutator between total density operators from being expressed as a linear combination of total density operators. 

\subsection{Chiral symmetry and interaction}
As shown in the previous section, generally, when the interaction is included, the ideal Euler bands are different from two copies of the ideal Chern bands. However, if the ideal Euler bands are the middle two bands in the system with chiral symmetry $S^2=1$, that anti-commutes with $I_{ST}^2=1$, then the Euler band can be decoupled into two Chern bands because $\rho_{+-}(\textbf{k})=\rho_{-+}(\textbf{k})=0$. In real materials, the chiral symmetry is nothing but the sublattice symmetry related to the invariance of the Hamiltonian under interchanging two sublattice sites. To prove it we use the form of Bloch wavefunction for middle two bands which are explained in the previous section. From Eq. (\ref{eq:middletwo}) in section \ref{chiralsection}, the Bloch wavefunctions for middle two bands can be written as
\begin{align}
    \ket{u_1(\textbf{k})}=
    \begin{pmatrix}
    \Re[\ket{\phi(\textbf{k})}]\\
    \Im[\ket{\phi(\textbf{k})}]
    \end{pmatrix},
    \ket{u_2(\textbf{k})}=
    \begin{pmatrix}
    \Re[i\ket{\phi(\textbf{k})}]\\
    \Im[i\ket{\phi(\textbf{k})}]
    \end{pmatrix},
\end{align}
where $\ket{\phi(\textbf{k})}$ is a complex vector, when $I_{ST}=K$, $S=\tau_y$, where $\tau_{x,y,z}$ indicate the sublattice degrees of freedom. Therefore, the Chern basis becomes,
\begin{align}
    \ket{u_-(\textbf{k})}=\frac{1}{\sqrt{2}}
    \begin{pmatrix}
    \ket{\phi(\textbf{k})}\\
    -i\ket{\phi(\textbf{k})}
    \end{pmatrix},
    \ket{u_+(\textbf{k})}=\left(\ket{u_-(\textbf{k})}\right)^*.
\end{align}
By unitary transform
\begin{align}
    U=\frac{1}{\sqrt{2}}
    \begin{pmatrix}
        I & -iI\\
        I & iI
    \end{pmatrix},
\end{align}
where $I$ is the identity matrix, the symmetry operators become
\begin{align}
    I_{ST}=\tau_xK,\quad S=\tau_z=
    \begin{pmatrix}
        I & 0\\
        0 & -I
    \end{pmatrix}
\end{align}
and the Bloch wavefunction becomes
\begin{align}
\label{eq:cherndensity}
    \ket{u_-(\textbf{k})}=
    \begin{pmatrix}
    0\\
    \ket{\phi(\textbf{k})}
    \end{pmatrix},
    \ket{u_+(\textbf{k})}=
    \begin{pmatrix}
    \left(\ket{\phi(\textbf{k})}\right)^*\\
    0
    \end{pmatrix}.
\end{align}
With the sublattice symmetry, orbitals are eigenstates of the chiral symmetry operator $S$. Therefore, because the orbitals are eigenstates for the density operator $e^{i\textbf{k}\cdot\textbf{r}}$, the density operator does not couple states with different eigenvalues for $S$. Consequently, since $\ket{u_+\left(\textbf{q}\right)}$, $\ket{u_-\left(\textbf{q}'\right)}$ have different eigenvalue of $S$, the density operator does not couple $\ket{u_+\left(\textbf{q}\right)}$, $\ket{u_-\left(\textbf{q}'\right)}$.
\begin{align}
    \bra{u_+\left(\textbf{q}\right)}e^{i\textbf{k}\cdot\textbf{r}}\ket{u_-\left(\textbf{q}'\right)}=0.
\end{align}
Now we can prove that
\begin{align}
\begin{split}
&\rho_{+-}(\textbf{k})\\
&=\int d\textbf{q}d\textbf{q}'
\ket{u_+\left(\textbf{q}\right)}\bra{u_+\left(\textbf{q}\right)}e^{i\textbf{k}\cdot\textbf{r}}\ket{u_-\left(\textbf{q}'\right)}\bra{u_-\left(\textbf{q}'\right)}=0,
\end{split}
\end{align}
\begin{align}
\begin{split}
&\rho_{-+}(\textbf{k})\\
&=\int d\textbf{q}d\textbf{q}'
\ket{u_-\left(\textbf{q}\right)}\bra{u_-\left(\textbf{q}\right)}e^{i\textbf{k}\cdot\textbf{r}}\ket{u_+\left(\textbf{q}'\right)}\bra{u_+\left(\textbf{q}'\right)}=0.
\end{split}
\end{align}
This allows separating the density operator into two parts, one projected to $\ket{u_+(\textbf{k})}$ and the other projected to $\ket{u_-(\textbf{k})}$.
\begin{align}
    \rho(\textbf{k})=\rho_{--}(\textbf{k})+\rho_{++}(\textbf{k}).
\end{align}
Therefore, the interacting physics for the Euler bands becomes the same as the interacting physics for each Chern band. Because of this in~\cite{Ideal_one}, they could have calculated the spectrum of chiral TBG with interaction by adding interaction to one of the two Chern bases instead of adding to two Euler bands.

\section{TBG continuum model}
The TBG Hamiltonian consists of the top layer term, the bottom layer term, and the tunneling term. 
\begin{align}
    &H_{TBG}=H_{+}+H_{-}+H_{T},\\
    &H_{+}=-t\sum_{\left<ij\right>}(c^{\dagger}_{iA+}c_{jB+}+c^{\dagger}_{jB+}c_{iA+}),\\
    &H_{-}=-t\sum_{\left<ij\right>}(c^{\dagger}_{iA-}c_{jB-}+c^{\dagger}_{jB-}c_{iA-}),\\
    &H_{T}=\sum_{i,j,\beta,\beta',l\neq l'}v(|\textbf{r}_{i\beta' l'}-\textbf{r}_{j\beta l}|)c^{\dagger}_{i\beta' l'}c_{j\beta l}.
\end{align}
where $+,-$ each denote top and bottom layer and $\beta=A,B$ denotes A or B site in graphene. Since graphene has half-filled Dirac cones at $\textbf{K}$ and $\textbf{K}'$ point, Hamiltonian represented in tight binding basis near $\textbf{K}$ point can be used to see the low energy physics of the TBG.
\begin{align}
\ket{\textbf{p},\beta,\pm}=\sum_{i}e^{i(\textbf{K}_\pm+\textbf{p})\cdot\textbf{r}_{i\beta \pm}}\ket{\textbf{r}_{i\beta \pm}}.
\end{align}
Here we note that $\textbf{K}$ point for the top layer and bottom layer is different because the layer is twisted. Under this basis, each layer of graphene can be approximated as a Dirac cone
\begin{align}
&\bra{\textbf{p}',\beta',l'}(H_++H_-)\ket{\textbf{p},\beta,l}=\delta_{l,l'}\delta_{\textbf{p}',\textbf{p}}h^l_{\beta'\beta}(\textbf{p}),\\
\label{eq:graphene}
&h^l_{\beta'\beta}(\textbf{p})=\begin{pmatrix}
    0 & (p_x+ip_y)e^{-il\theta/2}\\
    (p_x-ip_y)e^{il\theta/2} & 0\\
\end{pmatrix},
\end{align}
and the tunneling term has the form,
\begin{align}
&\bra{\textbf{p}',\beta',l'}H_T\ket{\textbf{p},\beta,l}=\omega \sum_{j=1}^{3} T^{\beta'\beta}_{j}\delta_{\textbf{p}',\textbf{p}-l\textbf{q}_j}(1-\delta_{l',l}),\\
&T_{j+1}^{\beta'\beta}=\omega(\sigma_0+\sigma_x \cos(j\phi)+\sigma_y \sin(j\phi)),\\
&\textbf{q}_1=\textbf{K}_--\textbf{K}_+, \textbf{q}_2=C_{3z}\left(\textbf{K}_--\textbf{K}_+\right),
\textbf{q}_3=C_{3z}^2\left(\textbf{K}_--\textbf{K}_+\right),
\end{align}
where $\phi=2\pi/3$.
To account for lattice relaxation, we reduce the size of the tunneling term related to repulsion between AA sites or BB sites, i.e. tunneling term for $\beta=\beta'$. The resulting tunneling term becomes,

\begin{align}\label{eq:tunneling}
&T_{j+1}^{\beta'\beta}=\omega_0\sigma_0+\omega_1(\sigma_x \cos(j\phi)+\sigma_y \sin(j\phi)).
\end{align}
where $0\leq\omega_0\leq\omega_1$.
In the main text, this Hamiltonian is represented in the real space, which is given by, 
\begin{align}
	H_{TBG}=
	\begin{pmatrix}
		-iv_0\boldsymbol{\sigma}_{\theta/2}\cdot\nabla & T(\textbf{r}) \\
		T^{\dagger}(\textbf{r}) & -iv_0\boldsymbol{\sigma}_{-\theta/2}\cdot\nabla  \\
	\end{pmatrix}
	,
\end{align}
where $T^{\beta'\beta}(\textbf{r})=\sum_{j=0}^2T^{\beta'\beta}_je^{i\textbf{q}_j\cdot\textbf{r}}$, $\nabla=\left(\partial_x,\partial_y\right)$, $\boldsymbol{\sigma}_{\theta/2}=e^{-i\theta\sigma_z/4}(\sigma_x,\sigma_y)e^{i\theta\sigma_z/4}$ with Pauli matrices $\sigma_{x,y,z}$ describing sublattice degrees of freedom in each graphene layer. In this form, it is easy to see that when $\omega_0=0$, the Bloch Hamiltonian has chiral symmetry
\begin{align}
	\begin{pmatrix}
        \sigma_z & 0\\
        0 & \sigma_z\\
	\end{pmatrix}.
\end{align}

To calculate the Euler curvature, a real Hamiltonian is needed. Real Hamiltonian can be obtained by a basis transform
\begin{align}
\ket{\textbf{p},a,l}=\frac{1}{\sqrt{2}}\left(\ket{\textbf{p},A,l}-i\ket{\textbf{p},B,l}\right),
\end{align}
\begin{align}
\ket{\textbf{p},b,l}=\frac{1}{\sqrt{2}}\left(-i\ket{\textbf{p},A,l}+\ket{\textbf{p},B,l}\right).
\end{align}
By this transform term from each layer, Eq.~(\ref{eq:graphene}) transforms to
\begin{align}
\begin{split}
h^l_{\beta'\beta}(\textbf{p})&=(p_x\cos\left(l\theta/2\right)+p_y\sin\left(l\theta/2\right))\sigma_x\\
&+(p_x\sin\left(l\theta/2\right)-p_y\cos\left(l\theta/2\right))\sigma_z.
\end{split}
\end{align}

and tunneling term Eq.~(\ref{eq:tunneling}) transforms to 
\begin{align}
&T_{j+1}^{\beta'\beta}=\omega_0\sigma_0+\omega_1(\sigma_x \cos(j\phi)+\sigma_z\sin(j\phi)).
\end{align}
lastly, the chiral symmetry operator transforms into $\sigma_y$.
For numerical calculation, a cutoff for $\textbf{p}$ is needed. The cutoff is applied by limiting the number of coupled k points. For this paper, we coupled up to the 11th nearest neighbor.

\begin{figure}
	\includegraphics[width=0.47\textwidth]{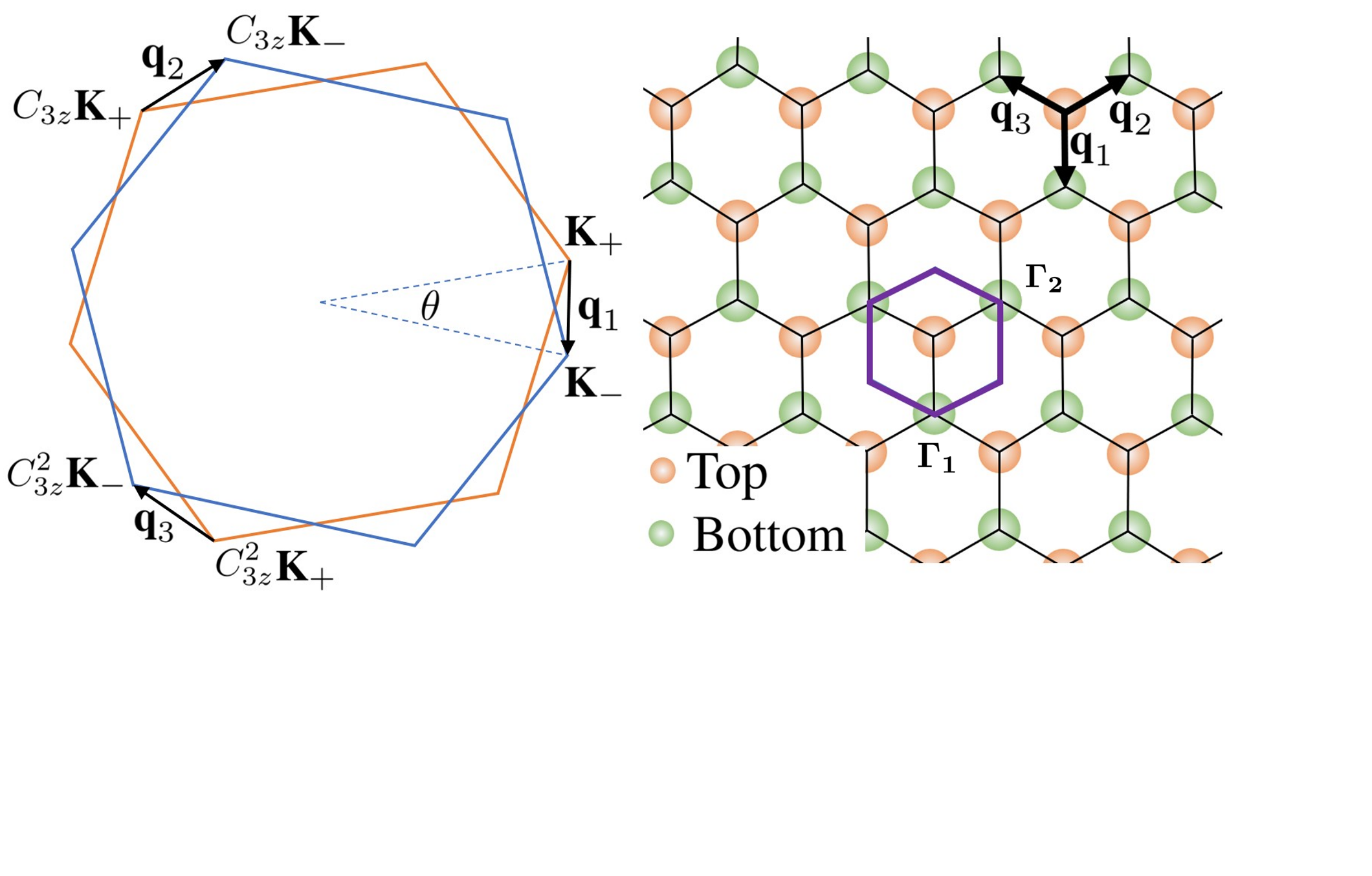}
  \vspace{-0.3cm}
	\caption{States coupled by tunneling term. The purple hexagon denotes the mBZ of the twisted bilayer graphene.}
 \label{fig:mBZ}
\end{figure}

Fig.~\ref{fig:mBZ}  shows how $q_j$ determines states that are coupled by the tunneling. The mini Brillouin zone(mBZ) of TBG is defined to have the same periodicity as the lattice in Fig.~\ref{fig:mBZ}. However, the system does not have the periodicity determined by mBZ. If the system has periodicity, the coupled states at $\Gamma_1$ and $\Gamma_2$ points should be the same. As we move from $\Gamma_1$ to $\Gamma_2$, the coupled state moves in an upright direction. As a result, the states in the left down direction are removed and the states in the right up direction are added to the list of coupled states. Due to the Dirac cone part of the Hamiltonian, removed and added states are related to the high energy physics of TBG because they are far from $\textbf{k}=0$. Therefore, TBG is not periodic in the energy scale near the cutoff. Physically, the high energy part of TBG is not periodic because of the quasi-periodicity of the exact structure of the twisted bilayer graphene. Only low-energy physics has periodicity. This allows the middle two bands in the TBG while having chiral symmetry $S^2=1$, that anti-commutes with $I_{ST}$.

\clearpage

\bibliography{merged}

\end{document}